\documentclass[aip,cha]{revtex4-1}

\usepackage{graphicx}
\usepackage{amsmath}
\usepackage{amsthm}
\usepackage{subfigure}
\usepackage{bm}
\usepackage{url}
\usepackage{appendix}
\usepackage[section]{placeins}
\renewcommand{\vec}[1]{\mathbf{#1}}

\usepackage{epsf}
\usepackage{epstopdf}

\newcommand{\pd}{\partial}
\newcommand{\ltn}{\lambda_{\text{TN}}}
\newcommand{\lf}{\lambda_{f}}
\newcommand{\h}{h}
\newcommand{\hf}{h_{f}}
\newcommand{\htn}{h_{\text{TN}}}

\newcommand{\tf}{\tau_{f}}

\newcommand{\pa}{PA}
\newcommand{\s}[1]{\sigma_{#1}}

\newcommand{\TNrep}{\Phi}

\begin{document}

\title{Topological chaos, braiding and bifurcation of almost-cyclic sets}

\author{Piyush Grover}
\affiliation{Mitsubishi Electric Research Laboratories, Cambridge, MA 02139, USA}
\author{Shane D. Ross}
\affiliation{Engineering Science and Mechanics, Virginia Tech, Blacksburg, VA 24061, USA}
\author{Mark A. Stremler}
\affiliation{Engineering Science and Mechanics, Virginia Tech, Blacksburg, VA 24061, USA}
\author{Pankaj Kumar}
\affiliation{Aerospace and Ocean Engineering, Virginia Tech, Blacksburg, VA 24061, USA}


\begin{abstract}
In certain two-dimensional time-dependent flows, the braiding of periodic orbits provides a way to analyze chaos in the system through application of the Thurston-Nielsen classification theorem (TNCT). We expand upon earlier work 
that introduced the application of the TNCT to braiding of almost-cyclic sets, which are individual components of almost-invariant sets [Stremler, Ross, Grover, Kumar, Topological chaos and periodic braiding of almost-cyclic sets. Physical Review Letters 106 (2011), 114101]. In this context, almost-cyclic sets are periodic regions in the flow with high local residence time that act as stirrers or `ghost rods' around which the surrounding fluid appears to be stretched and folded. 
In the present work, we discuss the bifurcation of the almost-cyclic sets as a system parameter is varied, which results in a sequence of topologically distinct braids. We show that, for Stokes' flow in a lid-driven cavity, these various braids give good lower bounds on the topological entropy over the respective parameter regimes in which they exist. We make the case that a topological analysis based on spatiotemporal braiding of almost-cyclic sets can be used for analyzing chaos in fluid flows. Hence we further develop a connection between  set-oriented statistical methods and topological methods, which promises to be an  
important analysis tool in the study of complex systems.
\end{abstract}
\maketitle

\pagenumbering{arabic}

\begin{quotation}
When a body of fluid moves, whether it be in the atmosphere, an ocean, or a kitchen sink, there are often regions of fluid that move together for an extended period of time.  As these 
coherent
 sets of fluid trace out trajectories in space and time, they can be thought of as `stirring'  the surrounding fluid.  The entanglement of these trajectories as they `braid' around each other is connected to the level of chaos present in the fluid system. 
When the sets return periodically to their initial positions in a  two-dimensional, time-dependent system, the entanglement of their trajectories can be used to predict a lower bound on the rate of stretching in the surrounding domain.   
We examine the trajectories of such `almost-cyclic sets' in a lid-driven cavity flow and demonstrate that this combination of topological analysis with set-oriented methods can be an effective means of predicting chaos.
The characterization of the entanglement and associated prediction of stretching are achieved through application of the Thurston--Nielsen Classification Theorem, which in general classifies the topological complexity of homeomorphisms of punctured surfaces. While a rigorous lower bound on topological entropy is not available in the absence of exactly periodic braiding structures, our approach finds the `topological skeleton' that can be used to get an approximate rate of stretching.
\end{quotation}

\section{Introduction} \label{mix_intro}

Qualitative and quantitative analyses of mixing in phase space are often key components of understanding the dynamics of a complex system.   Our focus is on systems exhibiting deterministic behavior, 
although our discussion is applicable to stochastic systems for which the full dynamics can be considered  a deterministic `template' on which an added stochastic behavior plays a secondary role.  
From the perspective of dynamical systems theory, the presence of mixing in phase space is synonymous with the existence of chaotic trajectories.  
There has been a significant amount of interest in understanding how such chaotic behavior arises, how to detect it, and how to design for or against it \cite{ott2002book,wiggins2003book}.

The field of fluid mechanics has been a particularly productive proving ground for understanding mixing in dynamical systems. 
Consider, for example, the development \cite{Aref1984} and growth \cite{aref2002pf} of 
the field of
chaotic advection, 
the phenomenon
in which advective transport by a regular velocity field leads to chaotic particle trajectories.  
Given an incompressible fluid moving with a velocity field $\vec{V} = (u,v,w)$, one can write the advection equations of motion for a passive fluid particle at position $\vec{x}(t)$ as
\begin{equation}
	\frac{\mathrm{d}\vec{x}}{\mathrm{d}t} = \vec{V}(\vec{x},t).
\end{equation}
Non-autonomous two-dimensional systems or autonomous three-dimensional systems are capable of generating chaotic particle trajectories even if the underlying velocity field is not chaotic or stochastic.  In the dynamical systems vernacular, a mixing system is one that generates chaotic trajectories.  From the perspective of fluid mechanics, this advective transport is typically referred to as `stirring', with the term `mixing' usually reserved for homogenization due to the combined effect of stirring and molecular diffusion.  It is usually assumed, and often correctly so, that dynamical systems mixing is synonymous with fluid mixing, and here we will use the terms interchangeably.
This approach to understanding and predicting  mixing has played an important role in the engineering analysis of laminar flows, and it has been  particularly valuable in the development of microfluidics \cite{ottino2004ptrsla,StrHasAref04}.
From the perspective of dynamical systems theory, this connection between regular flows and chaotic trajectories was previously known  \cite{arnold1968book}. However, it was not until the concept of chaotic advection took root that the general scientific community  fully realized the ubiquitous existence of relatively simple flows capable of generating chaotic transport.

In a typical analysis of  chaotic advection, one requires a detailed spatio-temporal  characterization of transport throughout the flow domain in order to quantify mixing
from the perspective of ergodic theory using metric quantities.  
Poincar\'{e} sections and Lyapunov exponents, 
and the closely-related measure-theoretic or Kolmogorov-Sinai (KS) entropy \cite{kolmogorov1958dras,sinai1959dras}, are generated by   
accurately tracking individual particles for very long times~\cite{wiggins2003book}.  Quantifications of transport using lobe dynamics involve tracking exponentially-stretched interfaces forward and backward in time in order to determine  stable and unstable manifolds 
and the area (or volume) that they encompass \cite{RoWi1990,Wiggins1992}. 
Lagrangian Coherent Structures (LCS), which can be viewed as generalizations of manifolds, are determined using the finite-time Lyapunov exponent field \cite{haller2001pd,haller2000c}.
Each of these methods provides important information about chaos in a domain, but they give information only in those regions of the domain where detailed 
metric calculations 
can be 
made, and having  incomplete information often prevents analysis.

In an alternative approach, mixing is quantified using topological, as opposed to metric, characteristics of the flow.   Applications of topology to analyzing fluid motion date back to Helmholtz \cite{helmholtz1858,ricca2009}, but the 
topological approach we employ here  is a modern development \cite{BoArSt2000}.  
The core idea in this approach is to use the topological classification of a few entangled periodic orbits to place a lower bound on the exponential growth rate of material surfaces under iteration of the flow.  
The connection between topology and exponential growth rates originated with
Adler, Konheim and McAndrew's definition of \emph{topological entropy}, which we will call $\h$, as an analogue of the KS entropy \cite{adler1965tams}.  
There is an enormous body of work concerning the dynamical meaning of $\h$ and ways to compute and estimate it.  Here we focus our 
attention on the main ideas and results that are relevant to the problem at hand.

The first key idea is the relation between $\h$ and the exponential growth rates of various quantities induced by iteration of a function.  
Perhaps the strongest such result is the proof by Yomdin \cite{yomdin1987ijm} of Shub's entropy conjecture that $\h$ is bounded below by the log of the spectral radius of the induced action on homology.  That is, the algebraic representation of the system topology (i.e.,~homology) produces a matrix representation of the iterated flow map (i.e.,~the action), and the log of the leading eigenvalue gives a lower bound on $\h$ for that flow.  Newhouse extended this result to the scenario of interest here by showing that for $C^{\infty}$-diffeomorphisms (i.e., infinitely differentiable, invertible mappings) of surfaces, which includes all singularity-free deterministic fluid flows, $\h$ is equal to the maximal growth rate of smooth arcs under the action of the diffeomorphism.  He also showed the important result that $\h$ varies continuously under infinitely smooth perturbations of the action, i.e., in $C^{\infty}$-families \cite{newhouse1989am,newhouse1991picm,NePi93}.

Also of importance here is the related (and earlier) result of Bowen \cite{bowen1978lnm} that $\h$ is bounded below by the growth rate of the induced action on the fundamental group, which consists of sets 
of equivalence classes of loops under homotopy, i.e.,~each set in the fundamental group consists of a class of loops that are equivalent under continuous deformation.  That is, Bowen proved that the growth rate of `representative loops' gives a lower bound on $\h$.  
That paper also contains what later became a key observation, namely, that by puncturing a surface at a periodic orbit and studying the action restricted to the punctured surface, one could get better lower bounds for the entropy using strictly algebraic means. Bowen\cite{bowen1978lnm} makes the prescient remark that Thurston's classification of surface isotopy classes can be used to good effect on the punctured surface.  

We thus come to the second key mathematical tool used here: Thurston's theorem on a surface punctured at a periodic orbit \cite{Thu88,casson1988book, fathi1991french, fathi2012english}.  Again, this concept has been the subject of a large body of work, which we consider here to only a limited extent; See Boyland\cite{boyland1994tia} for a excellent survey and additional references.  
The Thurston--Nielsen Classification Theorem (referred to hereafter as TNCT) characterizes the isotopy classes of 
continuous, one-to-one, invertible, orientable maps, or homeomorphisms, of punctured surfaces.  Since we are discussing the application of this work in the context of fluid flows, we will further restrict our discussion to differentiable maps of two-dimensional manifolds, i.e.~to surface diffeomorphisms.
The topological complexity in the system is introduced through the relative motions of the punctures induced by the flow map. The surfaces under consideration are $k$-punctured disks, $D_k$, in which the $k$ punctures can correspond to physical obstructions \cite{BoArSt2000,finn2003jfm}, periodic orbits \cite{GoThFi06,StrJie07}, or other stirrer-like objects that entangle the surrounding fluid, such as point vortices \cite{BoStAr2003}.  
 Since it is topology being considered here, the specific shapes of the disk and the punctures do not matter.
 In dynamical systems applications, these surfaces are typically two-dimensional physical space and the maps act in time, but under certain conditions maps can also be constructed in three-dimensional space \cite{JiSt09}.

An isotopy class consists of all the topologically equivalent diffeomorphisms that can be mapped to each other through a `fixed stirrer' diffeomorphism in which the punctures are fixed relative to each other during the mapping.  
According to the TNCT, each isotopy class contains a representative homeomorphism $\TNrep$, the \emph{Thurston--Nielsen (TN) representative}, that is one of three possible types: \emph{finite order} (FO), \emph{pseudo-Anosov} (\pa), or \emph{reducible}.  
%
If $\TNrep$ is FO, 
then its $n^{\text{th}}$ iterate is the identity map in which the  punctures are held fixed and all points in $D_k$ map to themselves, so no net stretching occurs.  All fixed stirrer diffeomorphisms are isotopic to the identity. 
If $\TNrep$ is PA, then it stretches and contracts everywhere on the surface by  factors $\ltn>1$ and $\ltn^{-1}$.   Iteration of a PA $\TNrep$ results in exponential stretching of smooth arcs with corresponding topological entropy $\htn = \ln\ltn$. 
There are several methods by which one can compute $\ltn$; for our analysis, we make use of the open-source code written by Hall \cite{toby2001}, which implements the Bestvina-Handel algorithm \cite{bestvina1995t}.    
A necessary (but not sufficient) condition that an isotopy class contain a \pa\ $\TNrep$ is the existence of at least three punctures in the surface. Finally, for the reducible case, $\TNrep$ leaves a family of curves invariant, and these curves delimit regions in which  $\TNrep$ is either FO or PA.   Each isotopy class contains only one TN representative, so the isotopy class itself can be referred to as being of FO, PA, or reducible type.  We focus here on  cases for which the isotopy class is of PA type.   

The power of Thurston's theorem in the analysis of dynamical systems comes from the connection between the TN representative  $\TNrep$ and the complexity of a flow $f$ within the same isotopy class.  By Handel's isotopy stability theorem  \cite{handel1985etds}, if $\TNrep$ is PA and $f$ is isotopic to $\TNrep$, then there exists a compact, $f$-invariant set $Y  \subseteq D_k$ and a continuous, onto mapping $\alpha : Y \rightarrow D_k$ such that $\alpha f = \TNrep\, \alpha$.  Thus, the complex dynamics of a PA $\TNrep$ are preserved by $f$ in (some subset of) the domain.  The map $\alpha$ may be many-to-one, so that the dynamics of $f$ may be more complicated than those of $\TNrep$, but they are never less complicated.  Therefore, the existence of a PA $\TNrep$ places a lower bound on the complexity of any flow map $f$ from the same isotopy class.  Non-trivial material lines, such as curves that encircle two punctures or that connect punctures with each other and/or the disk boundary, will grow exponentially in length under $n$ iterates of $f$ according to $L\sim{\lf} ^n$, with associated topological entropy  $\hf = \ln \lf$, and the TNCT establishes that $\lf \ge \ltn$ (and $\hf \ge \htn$).
Unfortunately, there are no restrictions on the size of $Y$, so the complicated dynamics may exist on a subset  with Lebesgue measure zero.  However,  the complexity associated with the  exponential stretching and folding of material lines cannot be removed by continuous perturbation of the domain that  maintains the orbit  topology. 
This topological chaos is thus ``built in'' to the system due to the orbit topology of only a finite number of punctures.

The initial application of the TNCT to fluid mixing examined a system of  three cylindrical  rods moving slowly on periodic  trajectories through a viscous fluid in a cylindrical domain \cite{BoArSt2000}.  The three rods exchanged positions pair-wise on circular trajectories.  
Two stirring protocols were considered: one motion  of FO type and one of \pa\ type. In the FO protocol, all motions involved clockwise interchanges.  In the \pa\ protocol, the exchanges alternated between clockwise and counterclockwise, producing figure-eight rod trajectories.  
 Although these two motions were energetically equivalent (in the Stokes' flow limit), the mixing produced by the \pa\ motion was clearly better than that produced by the FO motion.  Computational analysis of this flow system  \cite{finn2003jfm} confirmed that the lower bound predicted by the TNCT gives an excellent representation of the actual stretching rate caused by the \pa\ motion.  Furthermore, a substantial subset of the domain exhibited exponential stretching and folding, demonstrating that the dynamics of the TN representative can indeed be relevant to the analysis of a realistic fluid system.

Artin's braid group  \cite{Ma74} provides a useful framework for representing the trajectories of  
punctures under the action of the flow.
Each of the 
puncture trajectories is identified with a strand in a ``physical braid''.  
The temporal reordering of these strands  dictates the topological structure
 of the physical braid. 
The generator $\s{i}$ represents a clockwise exchange of strands $i$ and $i+1$, and $\s{-i}$ (our shorthand notation for $\s{i}^{-1}$) represents a counterclockwise exchange. 
A braid can be 
identified via a ``braid word'' of generator ``letters'', which we read from 
right to left as time progresses. 
For example, the braid on three strands that we discuss in Section~\ref{flow_description} is represented by the braid word $\s{-1}\s{2}$. 
Since the braid word, or the ``mathematical braid'', encodes only the topology of the braid crossings, each isotopy class on a punctured orientable two-dimensional surface is associated with one such braid.  Thus, we can refer to a braid as being of FO, PA, or reducible type.
A brief overview of the connection between braid theory and topological chaos is given in Section~\ref{reference_case}, and a general introduction to braid theory can be found in, e.g., \cite{birman1975,bangert2009}.

From the mathematical point of view it is clear that  the topological analysis of a fluid system can be based on the trajectories of periodic orbits, even when these orbits are not associated directly with stirring rods.  Such orbits can still appear to be `stirring' the surrounding fluid, and they have been thus been termed \emph{ghost rods} \cite{GoThFi06}.  In many cases, consideration of ghost rods is essential to understanding the presence of topological chaos  in a system.  For example, a single physical rod moving on an epicyclic trajectory in a two-dimensional domain \cite{GoThFi06} produces a braid on one strand, which is in an isotopy class of FO type. 
However, it is clear from direct observation that stirring a Stokes' flow with this motion produces  significant stretching and folding of material lines. This complexity can be explained using the TNCT by noting that there are two important fixed points, or ghost rods,  in the domain about which the moving rod winds.  
The braid on three strands corresponding to the space-time trajectories of the two fixed ghost rods \emph{and} the moving rod is in an isotopy class of \pa-type, and thus predicts the presence of chaos in the domain. 
The trajectories of moving ghost rods can also produce topological chaos, even when there are no physical rods in the system, such as in the lid-driven cavity flow discussed in Section~\ref{flow_description}
that forms the basis for our present discussion.  

Identifying 
braiding of ghost rods  is an important step in applying the concepts of topological chaos to a broad range of  fluid systems with no physical stirring rods.  However, finding  appropriate periodic orbits is a challenging task. One way of getting around this difficulty is to calculate the long-time braids generated by random collections of initial points \cite{Th05,Th10}.  The topological entropy is estimated from these computed braid generators.
This approach removes the need to identify periodic orbits in the flow but, since the braids being considered here are aperiodic, the TNCT cannot be invoked in the prediction of a lower bound.  Instead, relatively long-time calculations are needed.  It has also been found that a random selection of trajectories in the flow typically leads to a poor estimate of overall system behavior \cite{Th05},  so a large statistical representation is required in this approach.  



We have recently considered a different approach to examining topological chaos in flows without physical rods and with no 
discernible
 low-order periodic orbits on which to base a topological analysis  \cite{StRoGrKu2011}.  In this view, the domain is divided into distinct subsets such that there is a very small probability that typical trajectories beginning in each subset will leave this subset in a short time.  These \emph{Almost-Invariant Sets} (AISs) \cite{Dellnitz98onthe} can be determined from the eigenspectrum of the discretized Perron-Frobenius transfer operator via a set-oriented approach, which we review in Section~\ref{s:AIS}.  In some cases, disconnected components of an AIS correspond to almost-periodic regions, or what has been referred to as \emph{Almost-Cyclic Sets} (ACSs) \cite{Dellnitz98onthe}.  
Since the ACSs consist of trajectories that move together for a relatively long time, the region of the domain corresponding to an ACS can be identified as a (leaky) ghost rod, and a representative trajectory from each ACS can be used to reveal the underlying braid structure.  
Furthermore, since these ACSs contain almost-periodic orbits, the aperiodic braid they generate can be approximated by periodic continuation to a time-periodic braid.  
This time-periodic braid can be characterized using the TNCT, and an estimate of the topological entropy can hence be determined.  The topological entropy given by the TNCT is no longer a strict lower bound, since the time-periodic braid is only an approximation of the true dynamical structure.  However, it was demonstrated\cite{StRoGrKu2011} that this estimate can give a good representation of the flow.  

Our earlier work\cite{StRoGrKu2011} considers creeping flow in a lid-driven cavity and starts with parameters for which there exist three periodic orbits, which we refer to as the reference case.  For small perturbations from the reference case, consideration of ACS braiding gives a clear explanation for the topological entropy in this flow.  However, 
for certain larger perturbations from the reference case, we observe a drop in the numerically computed topological entropy below the lower bound predicted for the reference case.  
This paper extends the previous analysis to consider these larger perturbations,  which leads us to our main observation. As we vary 
away from the reference case, the AIS/ACS structure appears to bifurcate, leading to a sequence of topologically distinct braids with differing numbers of strands.  The motion in $D_k$ for each of these new braids on $k$ strands belongs to an isotopy class of PA type, and the corresponding value of $\ltn$   gives a correct lower bound estimate on the topological entropy for the corresponding parameter value.  Hence, we give further evidence that these almost-cyclic sets are  natural objects on which to base an application of the TNCT.
We assert that this work marks an important step in making the ghost rod methodology applicable to realistic fluid flows with arbitrary time-dependence when low-order periodic, braiding orbits are difficult to identify or do not  exist. We also conjecture that this generalization is applicable to a wide variety of dynamical systems,  not just fluid systems.


\section{Braiding of periodic orbits in a  lid-driven cavity flow}\label{flow_description}

\subsection{The reference case}\label{reference_case}

The fluid system model that we analyze in this work is a simplified version of a system\cite{MeGo2004note} that allows for exact solutions. This relatively simple fluid system has an easily-visualized piecewise-steady velocity field, but due to the imposed time dependence it exhibits complicated dynamics.

We consider a two-dimensional lid-driven cavity flow in an infinitely-wide cavity with height $2b$.  Under the assumption of Stokes' flow, the stream function $\psi(x,y)$ defined by
\begin{equation}\label{advection}
	\mathbf{V}(\mathbf{x},t) = \left( \frac{\pd\psi}{\pd y}, -\frac{\pd\psi}{\pd x}, 0 \right)
\end{equation} 
satisfies the two-dimensional biharmonic equation,
\begin{align}\label{eq:eom}
\nabla^2\nabla^2\psi(x,y)=0.
\end{align}
We assume that the flow is driven by prescribed tangential velocities on the top and bottom boundaries at $y=\pm b$.  The piecewise steady tangential velocities we take to be
\begin{equation}\label{eq:V}
\displaystyle
u(x,b) = -u(x,-b) =\begin{cases}
~~U_1\sin\left( \frac{\pi x}{a} \right)+U_2\sin\left( \frac{2\pi x}{a} \right) &\text{when $n \tau_f \leq t < (n+1) \tau_f/2$}\\
-U_1\sin\left( \frac{\pi x}{a} \right)+U_2\sin\left( \frac{2\pi x}{a} \right) &\text{when $ (n+1)\tau_f/2 \leq t <  (n+1) \tau_f$}
\end{cases}
\end{equation}
for integer $n$, where $\tau_f$ is the time period of the system.
The solution to Eq.~(\ref{eq:eom}) subject to the boundary conditions in Eq.~(\ref{eq:V}) is 
\begin{subequations}\label{eq:psi_solution}
\begin{equation}\label{eq:assume_stream_function}
\psi(x,y,t)=\begin{cases}
~~U_1C_{1}f_{1}(y)\sin\left(\frac{\pi x}{a}\right)+U_2C_{2}f_{2}(y)\sin\left(\frac{2\pi x}{a}\right) & n \tau_f \leq t < (n+1) \tau_f/2 \\
-U_1C_{1}f_{1}(y)\sin\left(\frac{\pi x}{a}\right)+U_2C_{2}f_{2}(y)\sin\left(\frac{2\pi x}{a}\right)&(n+1)\tau_f/2 \leq t <  (n+1) \tau_f,
\end{cases}
\end{equation} 
where
\begin{equation}\label{eq:f_function}
\begin{split}
f_{k}(y)=\frac{2\pi y}{a}\cosh\left(\frac{k\pi b}{a}\right)\sinh\left(\frac{k\pi y}{a}\right)-  \frac{2\pi b}{a}\sinh\left(\frac{k\pi b}{a}\right)\cosh\left(\frac{k\pi y}{a}\right)
\end{split}
\end{equation}
and
\begin{equation}\label{eq:C_function}
C_{k}=\frac{a^2}{2k\pi^2b}\left[\frac{a}{2k\pi b}\sinh\left(\frac{2k\pi b}{a}\right)+1\right]^{-1}.
\end{equation}
\end{subequations}
The spatial symmetry in the boundary conditions produces a vertical streamline in the flow at $x= n\,a$, and without any loss of generality we can restrict our attention to the bounded two-dimensional rectangular domain 
\begin{align}
D = \{(x, y) : 0\leq x \leq a,-b\leq y \leq b\}.
\end{align}
By the time periodicity of the boundary conditions, the flow pattern is reflected about the line $x=a/2$ every $\tau_f$/2 units of time.  

In Section~3 we will often refer to the one-parameter family of stroboscopic global Poincar\'e maps,
\begin{equation}\label{flowmap}
\phi_t^{t+\tau_f}:D \rightarrow D, 
\end{equation}
where $\phi_t^{t+\tau_f}$ is the solution diffeomorphism (i.e., flow map) from time $t$ to $t+\tau_f$; that is, $\phi_t^{t+\tau_f}$ represents the motion of all fluid particles over one period of the velocity field starting at time $t \in [0,\tau_f]$. Here we treat the initial time $t$ as a  parameter, which we can also think of as an initial phase (belonging to $S^1$).

\begin{figure}[b!]\centering
\includegraphics[width=16cm]{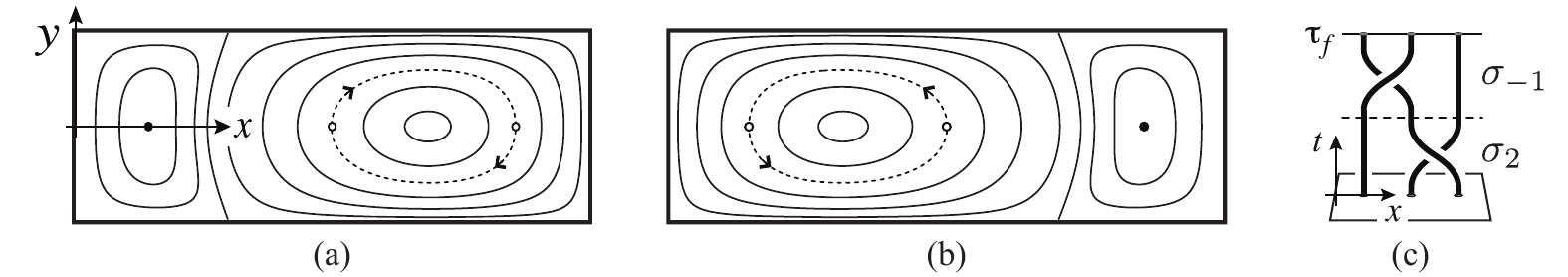}
\caption{
(a--b)~Representative streamlines given by Eq.~(\ref{eq:psi_solution}) for the reference case with $a/(2b) = 3$ and $U_2/U_1\approx 0.841$ for  (a)~$0\leq t < \tau_f/2$ and (b)~$\tau_f/2 \leq t < \tau_f$.  Points marked by solid circles are stagnation points in the flow.  If $a=6$ and $U_1 \approx 0.1055$, 
 the fluid particles starting at the open circles have exchanged positions after $\Delta t = \tau_f/2=1/2$.
(c)~The braid diagram generated by the periodic orbit trajectories in the reference case.
}
\label{cavity_flow_streamlines}
\end{figure}

We will focus on analyzing behavior in  a domain with aspect ratio $a/(2b)=3$, such
as is shown in figure \ref{cavity_flow_streamlines}. 
The structure of the flow 
for the reference case depends on the value of the  parameters $a$, $U_1$, $U_2$, and $\tau_f$.  The ratio $U_2/U_1$ dictates the streamline pattern, and the relationship between $U_1$ (say) and $\tau_f$ determines how much time a particle spends traveling along a streamline before the flow is `blinked' to another streamline pattern.  
For a domain aspect ratio of $a/(2b) = 3$, we have the following properties when we take $a=6$, $U_2/U_1\approx 0.841$, $U_1 \approx 0.1055$, and $\tau_f=1$: 
\begin{enumerate}
\renewcommand{\labelenumi}{$$\arabic{enumi}$)$}

	\item  There exist three points $\mathbf{x}_L$, $\mathbf{x}_C$ and $\mathbf{x}_R$, such that $\mathbf{x}_C$ is at the center of the domain $(x,y)=(a/2,0)$,  and $\mathbf{x_L}$ and $\mathbf{x}_R$ are located symmetrically about $x=a/2$ along $y=0$.

	\item  For $0\leq t<\tau_f/2$, $\mathbf{x}_L$ is a fixed point, and $\mathbf{x}_C$ and $\mathbf{x_R}$ exchange their positions while moving \textit{clockwise} along their shared streamline.

	\item Since the flow pattern is reflected about $x=a/2$ at $t=\tau_f/2$, $\mathbf{x}_R$ is a fixed point  for $\tau_f/2\leq t < \tau_f$, and $\mathbf{x_C}$ and $\mathbf{x}_L$ exchange their positions during this time period while moving \textit{counterclockwise}.

\end{enumerate}
After three periods of the flow, the  points  $\mathbf{x}_L$, $\mathbf{x}_C$ and $\mathbf{x}_R$ return to their original positions. 
We refer to this choice of  parameters as the \emph{reference case}, for which the trajectories of these three period-3 points form the (2+1)-dimensional space-time braid on three strands, $\s{-1}\s{2}$,  shown in figure \ref{cavity_flow_streamlines}(c).
We determine this braid structure by projecting the trajectory crossings onto the $x$-axis (which gives the `physical braid representation') and follow the braid group labeling convention discussed in Section~1.

\begin{figure}[b!]
\begin{center}
\includegraphics[width=16cm]{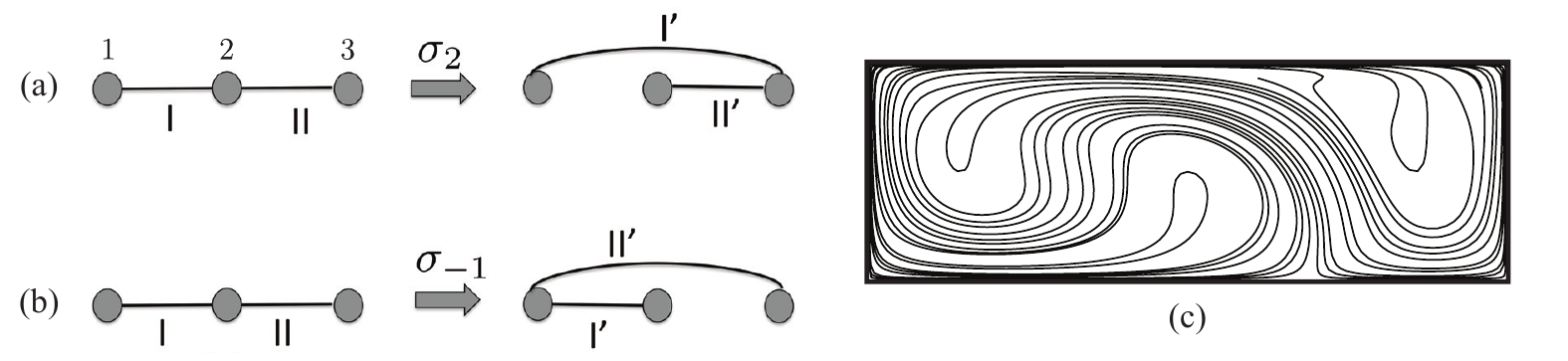}
\caption{The stretching of material lines under the action of the two generators in the reference case:  (a)~the action of $\sigma_2$, (b)~the action of $\sigma_{-1}$, and (c)~the actual  stretching of a material line that is initially along the $x$ axis in the reference case.}
\label{fig:linestretch}
\end{center}
\end{figure}

In the reference case, the determination of $\ltn$ can be described via the Burau representation \cite{burau1936}.   In this representation, a braid on $N$ strands is identified with a $(N-1) \times (N-1)$ matrix.  Each of the braid generators has a matrix representation, and the braid matrix is given by the product of the generator matrices. The entries of Burau matrices are Laurent polynomials in some variable $t$, which is taken to be $-1$ here.
The Burau matrices can be heuristically understood as Markov matrices that track the evolution of material lines connecting the points. For instance, in figure \ref{fig:linestretch}, the generator $\sigma_2$ stretches the line I to approximately look like the sum of lines I and II while leaving the line II intact, giving the relations I'$=$I+II and II'$=$II.  The Burau matrix corresponding to $\sigma_{2}$ is thus
\begin{equation}
M_2=\left( \begin{array}{cc}
1 & 0 \\
1 & 1 
 \end{array} \right).
 \end{equation}
Similarly, for $\sigma_{-1}$ the Burau matrix is,
\begin{equation}
M_{-1}=\left( \begin{array}{cc}
1 & 1 \\
0 & 1 
 \end{array} \right).
 \end{equation}
The Burau matrix for the braid $\s{-1}\s{2}$ is then $M = M_{-1}\,M_2$.  According to the TNCT, when $N=3$ the stretching rate produced by the TN representative $\TNrep$ 
 is the same as in the linear map represented by the Burau matrix.  That is, for the reference case $\ltn$ is given by the dominant eigenvalue of $M$, so that $\ltn =(3+\sqrt 5)/2$.  In order to distinguish between the stretching produced by topologically distinct flows with $N$ punctures, we will refer to the entropy in this reference case as  $h_{\text{TN},3} = \log(\ltn) \approx 0.962$.
 

For $N>3$, the dominant eigenvalue of the Burau matrix only provides a lower bound for $\ltn$, and other methods must be used instead, such as the train-tracks algorithm \cite{bestvina1995t, toby2001} or the encoding of loops  \cite{hall2009tia,Th10}.


The actual topological entropy produced by the flow map, $h_f$,  can be determined by computing the asymptotic stretching rate of topologically non-trivial lines \cite{newhouse1993a}, such as  lines that join a periodic point with the outer boundary. 
We estimate $h_f$ by following two orthogonal lines that are initially along the lines $x=a/2$ and $y=0$, respectively, as they are stretched over 6--10 periods of the flow.  For numerical reasons these lines are not extended all the way to the boundary, but instead are terminated a distance $\epsilon < a/200$ from the boundary.  
For this reference case, these computations give $\hf\approx 0.97$, which is reasonably well represented by the lower bound, $h_{\text{TN},3}$.



\subsection{Perturbation of the reference case}\label{sec:cases}

\begin{figure}[ht!]
\begin{center}
\includegraphics[width=0.5\textwidth]{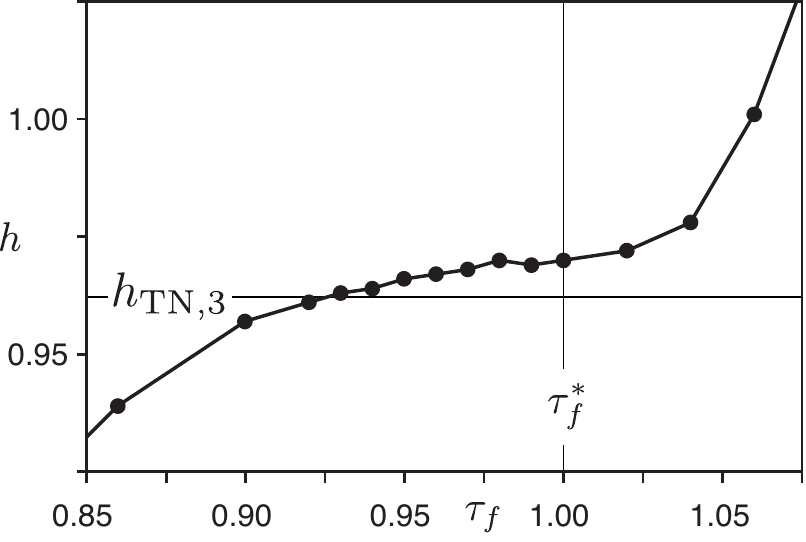}
\caption{\footnotesize {Variation of the actual topological entropy of the fluid, $h_f$, as a function of $\tau_f$. Also shown is the lower bound $h_{{\rm TN},3}$  for $\s{-1}\s{2}$,  the braid on 3 strands corresponding to the reference point with  $\tau_f = 1$.}}
\label{fig:entropy}
\end{center}
\end{figure}

For the analysis discussed here, we  keep the ratio $U_2/U_1$ fixed while changing the value of $\tau_f$ away from unity. 
This perturbation increases (for $\tf>1$) or decreases (for $\tf<1$) the amount of rotation in the domain before switching, which  has a direct effect on the stretching of material lines.  The computed topological entropy for the flow is shown in figure~\ref{fig:entropy} for a range of $\tf$ values.  
The entropy varies smoothly, as expected from the results of Newhouse \cite{newhouse1989am}.  The overall trend in $h_f$ is logical, as increasing $\tau_f$ increases the energy added to the flow during each period.  However, the rate of increase in $h_f$ is not uniform, which we explain by considering the topology of the flow map.



\begin{figure}[t]
\centering
\includegraphics[width=8cm]{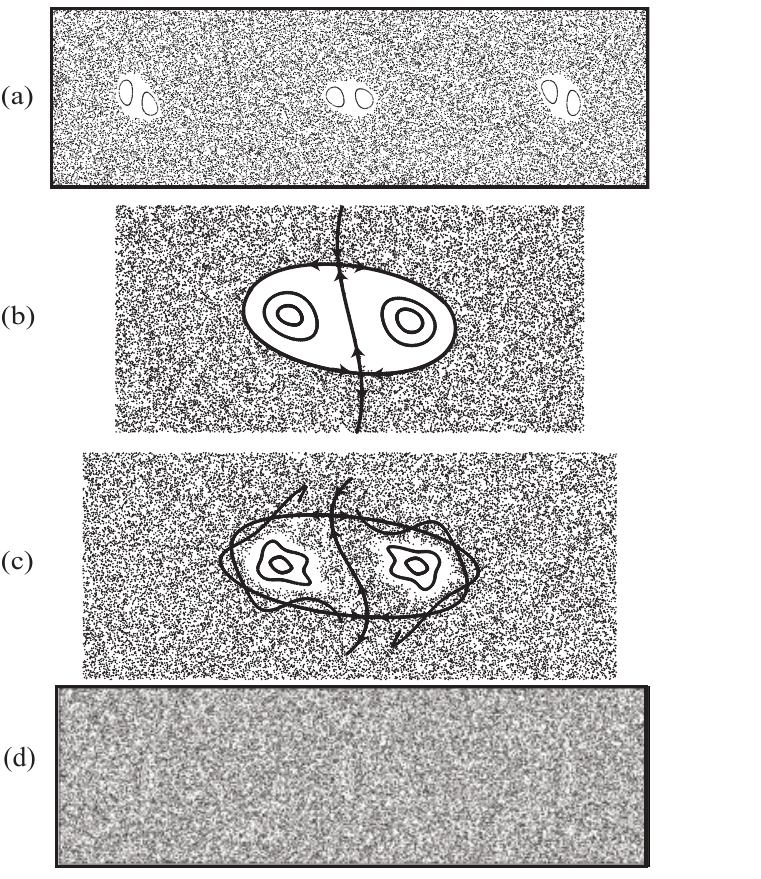}\\
\caption{\footnotesize{(a)~Poincar\'e section from the map $\phi_0^{\tf}$ for $\tau_f=1.01$. Note the  `leaky ghost rods' generated by perturbation of the parabolic periodic points from the reference case. (b, c)~Close-ups of the 
domain near  $(x,y)=($a/2$,0)$, which show the structure of the center `ghost rod' consisting of two saddle and two centers for (b)~$\tau_f=1.01$ and (c)~$\tau_f=1.04$. In (b) the stable and unstable manifolds nearly overlap, while in (c) they clearly intersect transversally, leading to significant `leaking' in and out of the region bounded by them. 
(d)~Poincar\'e section from the map $\phi_0^{\tf}$ for $\tau_f\approx0.99$. The phase space appears featureless and is devoid of low-order periodic orbits that form non-trivial braids.
}}
\label{fig:poin_101}
\end{figure}

In the reference case with $\tau_f=1$, the three period-3 points discussed in section~\ref{flow_description} are parabolic points, which are structurally unstable.  Under small perturbations for $\tf>1$, each parabolic  point bifurcates into a pair of periodic saddles and periodic centers, as illustrated in figure~\ref{fig:poin_101}.  The stable and unstable manifolds that pass through the saddles (i.e., the hyperbolic points) form barriers to transport, so that fluid within these regions moves together for a number of periods.  These `ghost rod' structures are `leaky', in that there is a small amount of transport across their boundaries due to the transverse intersections of the stable and unstable manifolds, form lobes. The amount of leakage per period is the amount of phase space enclosed by the lobe \cite{Wiggins1992}. In figure \ref{fig:poin_101}(b,c), we show the manifold structure for the center `leaky ghost rod' for $\tau_f=1.01$ (panel~b, where the the transversal intersection is present, but not evident) and $\tau_f=1.04$ (panel~c). The four period-3 points (two hyperbolic and two elliptic) that form the structure of a single `leaky ghost rod' produce a braid on 12 strands.  The four strands corresponding to each individual ghost rod structure simply twist around each other, and as a result the braid on 12 strands is reducible to a braid on 3 strands.  
This mathematical braid on 3 strands is identical to the mathematical braid from the reference case.  Thus, the value of $\htn\approx 0.962$ predicted by the TNCT for the reference case remains the lower bound on $h_f$  for the perturbed flow we have considered here with $\tf>1$. 
 Clearly there is additional stretching in these cases that is not captured by the TN representative, which suggests the presence of additional ghost rods. We do not explore this increased complexity here.

When the value of $\tau_f$ is decreased below 1, the parabolic points 
disappear and no low-order periodic points are found in the flow.  
We have searched numerically for periodic orbits by iteratively mapping regions of the flow forward and backward in time.  
For example, the Poincar\'e section shown in figure~\ref{fig:poin_101}(d) for $\tau_f\approx0.99$ appears to be completely chaotic.  Thus, there are no apparent periodic `ghost rods' available to form a braid and predict a lower bound on the topological entropy of the flow. However, figure~\ref{fig:entropy} shows that $h_{TN,3}$  is a lower bound for $h_f$ for approximately a $7\%$ perturbation in $\tau_f$\cite{StRoGrKu2011}. In the discussion below, we consider the validity of the topological entropy predicted by TNCT even if we cannot identify  exactly periodic orbits that produce the expected braiding motion. We accomplish this by using a transfer operator approach to reveal phase space structures that braid non-trivially and persist under perturbations in the $\tau_f$ values.
 


\section{Almost-invariant and almost-cyclic sets}\label{AISsec}

%
 
\subsection{Computation of almost-invariant and almost-cyclic sets}\label{s:AIS}

\begin{figure}[h!]
\centering
\includegraphics[width=2in]{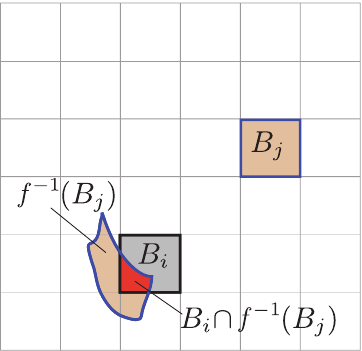}
\caption{\footnotesize{
Computation of the transition matrix $P_{t,\tau_f}$ by a set-oriented method. The phase space is divided into boxes, $\{B_1,B_2,\ldots,B_n\}$.
Box $B_j$ at  the final time $t+\tau_f$ is mapped (backwards) to $f^{-1}(B_j)$ at the initial time $t$, where $f=\phi_t^{t+\tau_f}$.
The value of the entry $p_{ij}$ is the fraction of box $B_i$ that is mapped into box $B_j$ by $f$. 
}}
\label{fig:AISillus}
\end{figure}

We use a set-oriented method to compute AISs for the lid-driven cavity flow system. Our aim is to partition the phase space into a given number (say $k$) of sets $\{A_1,A_2,....A_k\}$ such that the phase space transport between these sets is very unlikely. We also want these sets to be important statistically with respect to the long term dynamics of the system. Following \cite{FrPa09}, this can be formulated  as follows.
We define the invariance of a set $A_i$ as
\begin{align}\label{invariance}
\rho_\mu(A_i)=\frac{\mu(A_i \cap f^{-1}(A_i))}{\mu(A_i)}.
\end{align} 
This quantity represents the probability (according to an invariant measure $\mu$) of a point in set $A_i$ being mapped back to $A_i$ over one iteration of a map $f=\phi_t^{t+\tau_f}$. We want to maximize the quantity $\displaystyle \sum_{i=1}^k\rho_\mu(A_i)$, with the constraint that $\mu(A_i)$ is not too small compared to 1. Without loss of generality, we can restrict to  partitions $\{A_1,A_2,....A_k\}$ such that each member $A_j$ of 
is a union of  sets in $\{B_1,B_2,\ldots,B_n\}$, i.e., each $A_j= \cup_{i\in I}B_i$, for some set of box indices $I\subset\{1,2,....,n\}$. 
Using the Ulam-Galerkin method \cite{Ulam1964,Li1976,Bollt2012}, we define the \emph{transition matrix} $P_{t,\tau_f}$ 
with entries
\begin{align}\label{matrix_entries}
p_{ij}=\frac{m\left(f^{-1}(B_{i})\cap B_{j}\right)}{m(B_{j})},
\end{align} 
where $B_{1},\ldots,B_{n}$ are the boxes in the covering and $m$ is the normalized Lebesgue measure, which coincides with the phase space volume measure (see figure \ref{fig:AISillus}).
In our computations, all boxes will have the same measure, i.e., $m(B_{i})=m(B_{j})$ for all $i,j$.

%
This stochastic matrix $P_{t,\tau_f}$ is a discretized approximation of the Perron-Frobenius operator of the map $f=\phi_t^{t+\tau_f}$ 
that may be viewed as a transition matrix of an $n$-state Markov chain
\cite{Bollt2002}. 
The first left eigenvector $v_1$ of $P_{t,\tau_f}$, corresponding to an eigenvalue $\lambda_1=1$, is the invariant distribution of the system, i.e., the discretized version of the invariant measure $\mu$ \cite{Dellnitz98onthe}. 

Following \cite{Froyland2005}, we form a new reversible matrix $R_{t,\tau_f}$, which is a stochastic matrix
having only real eigenvalues \cite{Bremaud1999} that satisfies important properties related to almost-invariance \cite{Froyland2005}.
First, we define the \emph{reverse time} transition matrix $\hat{P}_{t,\tau_f}$, whose elements are given by
\[ 
\hat{p}_{ij}=\frac{v_{1,j}}{v_{1,i}}p_{ji}.
\]
Here, by $v_{1,i}$ we mean the $i$th component of the vector $v_1$, i.e., the value of $v_1$ assigned to the box $B_i$.
(For additional accuracy, one can directly calculate $\hat{P}_{t,\tau_f}$ as ${P}_{t,-\tau_f}$ \cite{FrPa09}.)
The reversible matrix $R_{t,\tau_f}$ is
\[
R_{t,\tau_f}=\tfrac{1}{2}(P_{t,\tau_f}+\hat{P}_{t,\tau_f}).
\]
We use the left eigenvectors $v_k$, corresponding to eigenvalues $\lambda_k$ of $R_{t,\tau_f}$, to detect AISs.
If an AIS (over one period-$\tau_f$ map, $\phi_t^{t+\tau_f}$) 
can be further decomposed into period-$N$ subsets, then those subsets are ACSs of the flow of period $N\tau_f$, and are hence AISs under the map $(\phi_t^{t+\tau_f})^{N}=\phi_t^{t+N\tau_f}$.

For this system, we approximate the dynamics on $D$ by covering it with a collection of $n=19200$ equally sized square boxes (aspect ratio 1), i.e., 240 boxes in the $x$ direction times 80 boxes in the $y$ direction. We take 100 uniformly distributed points in each box (a 10$\times$10 grid), perform a forward iteration for each point for one period of the flow $\tau_f$, and monitor the boxes reached by the iteration. The transition probability from a source box to a destination box is measured by the ratio of the number of points from the source box that reach the destination box in one iteration step.

All  subsequent computations were verified to depend weakly on the number of points per box 
(for values greater than 100) by comparing the results with those obtained when using 2025 points per box (a 45$\times$45 grid)
instead of 100, with the difference in the various eigenvalues for the two cases being less than $0.1\%$ for 
the range of parameters considered here. Similarly, a finer discretization of phase space, obtained by using 30000
boxes ($300 \times 100$ boxes) was used to verify the convergence of eigenvalues. It was also verified that finer discretization does not change the order of the eigenvectors.
 
The system being considered here is a canonical Hamiltonian system, and thus the invariant measure, $\mu$,
is the same as volume measure, $m$.
As our boxes are of equal sizes, the entries of $v_1$ should be approximately equal.
We note that the invariant measure is not unique in our case, since a point measure at the periodic points (if any) will also be invariant. However, due to the discretization procedure, we recover the unique absolutely continuous measure.

We use the sign structure of the $v_k$, $k\ge 2$, to detect AISs \cite{FrDe2003,FrPa09}.
For instance, for a given scalar value $c$ (chosen as described below), the set given by the union of regions in phase space such that the component of $v_2$ is greater than $c$ corresponds to one AIS, while
the union of regions such that the component of $v_2$ is less than $c$ corresponds to its complement, another AIS.  
More precisely, the two AISs are defined as follows. 
Let $I_1=\{i: v_{2,i}>c, 1\leq i \leq n\}$ and $I_2=\{i: v_{2,i}<c, 1\leq i \leq n\}$, then $A_1=\cup_{i\in I_1} B_i$ and $A_2=\cup_{i\in I_2} B_i$. The value of $c$ is chosen so as to maximize $\min(\rho_\mu(A_1),\rho_\mu(A_2))$. 
For the majority of cases, we will take $c=0$, but to maintain consistency across figures, for non-zero $c$ we plot $(v_{2,i}-c)$ values while identifying the AISs, so that the zero contour on the figure always refers to boundaries delineating the AISs.

It has been shown\cite{FrPa09} that if $\rho_\mu(A_m)\leq\rho_\mu(A_n)$, then $\mu(A_m)\leq\mu(A_n)$. It has also been shown \cite{Froyland2005} that for any $A_m$, such that $\mu(A_m)<\tfrac{1}{2}$, there exist the following bounds on transport,
\begin{equation}\label{bounds}
 \rho_\mu(A_m) \in \left( 1-\sqrt{2(1-\lambda_2)} , \tfrac{1}{2} (1+\lambda_2) \right).
\end{equation}

%

\begin{figure}[h!]
\includegraphics[width=.95\textwidth]{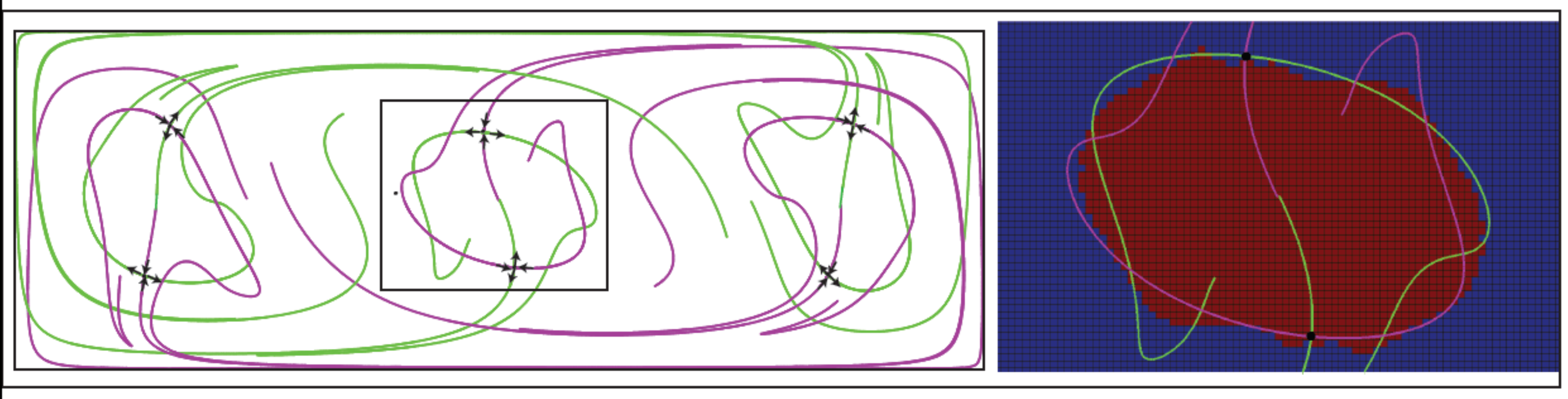}
\caption{\footnotesize (Left) The stable (magenta) and unstable (green) manifolds of the period-3 saddle points for $\tau_f\approx1.06$. (Right) The manifolds superimposed with a two-set decomposition of the phase space, shown near the point  $(x,y)=(3,0)$. }
\label{fig:lobe_dynamics}
\end{figure}

Before moving to an investigation with $\tau_f<1$, it is instructive to study the transport in our system for $\tau_f>1$, since we can compare the set-oriented method results with those obtained from lobe dynamics. In figure \ref{fig:lobe_dynamics}, we show the stable and unstable manifolds of the period-3 saddle points for the case $\tau_f\approx1.06$. 
On the right, we show the AIS decomposition into two sets (red and blue) near the point $(x,y)=(3,0)$. 
We are interested in finding the  
transport out of the red region under $f^3$, where $f=\phi_t^{t+\tau_f}$. For the 
transition matrix for $f$, 
$\lambda_2=0.9967$, and thus for $f^3$ we have $\lambda_2=(0.9967)^3=0.990$. 
Hence the lower  and upper bounds on invariance (\ref{bounds}) tell us that, under $f^3$, 
$\rho_\mu(A_1) \in (0.857, 0.995)$, 
where $A_1$ and $A_2$ are defined as before, and $\mu(A_1)<\tfrac{1}{2}$.
Lobe dynamics theory 
 \cite{RoWi1990, RoWi1991, DeJuKoLeLoMaPaPrRoTh2005, DuMa2010, RoTa2012}
 tells us that the amount of phase space transported out of a  separatrix-bounded region (boundaries given by stable and unstable manifolds from saddle to primary intersection points) is given by the size of the lobe. Using box counting, we find that one lobe consists of approximately 56 boxes. There are 2 lobes, one on each side, and hence the approximate transport out of the 
 separatrix-bounded region 
 under $f^3$ equals 112 boxes. 
Dividing this area by the size of the 
separatrix-bounded region 
(=1158 boxes), we find that the invariance predicted by lobe dynamics is $=1-112/1158 \approx 0.903$, well within the bounds provided by $\lambda_2$.

\begin{figure}[h!]
\centering
\includegraphics[height=1.5in,width=4.5in]{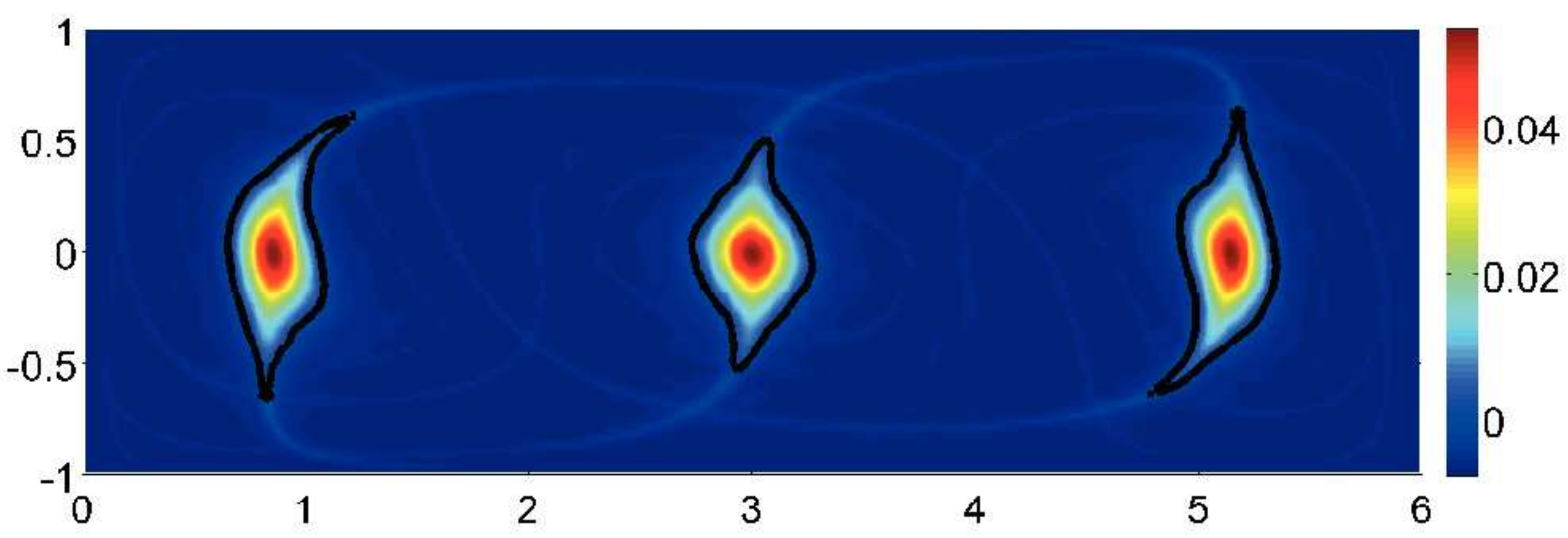}
\caption{\footnotesize {AIS structure for $\tau_f\approx0.99$ based on $v_2$. The zero contour (black) is the boundary between the two almost invariant sets. Compare with the Poincar\'e section for the same $\tau_f$ shown in figure \ref{fig:poin_101}(d)}}
\label{fig:tau_cric}
\end{figure}

\begin{figure}[h!]
\begin{center}
\subfigure[$v_3$]{
\includegraphics[width=3in, height=1in]{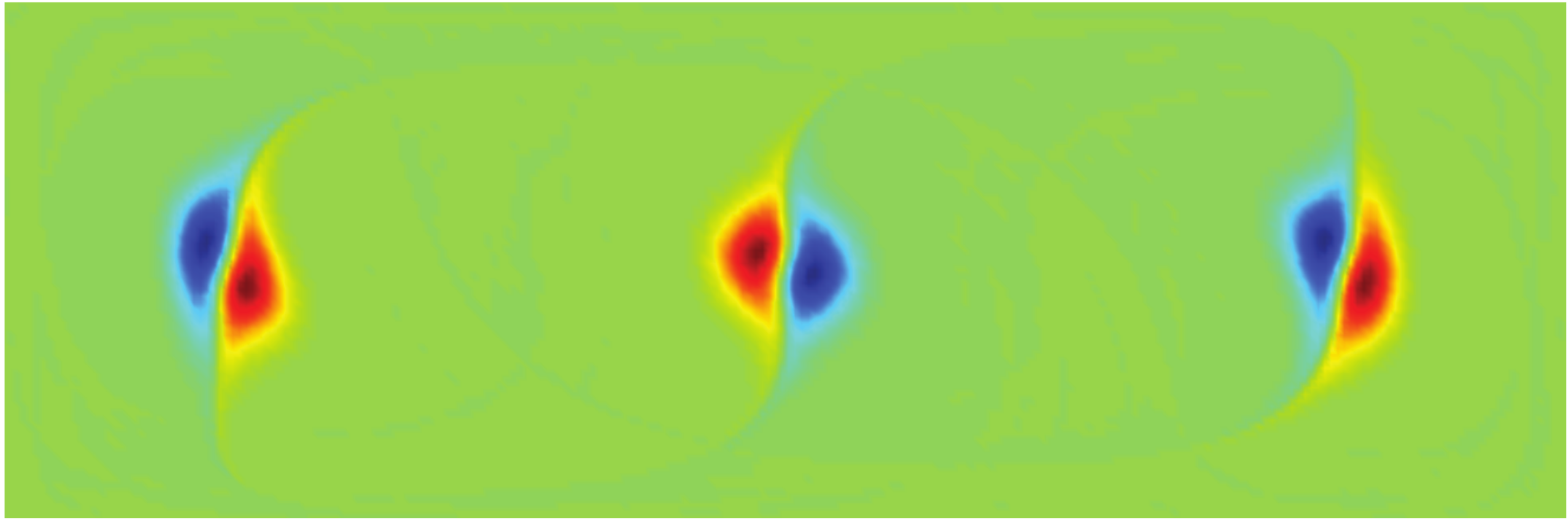}}
\subfigure[$v_4$]{
\includegraphics[width=3in, height=1in]{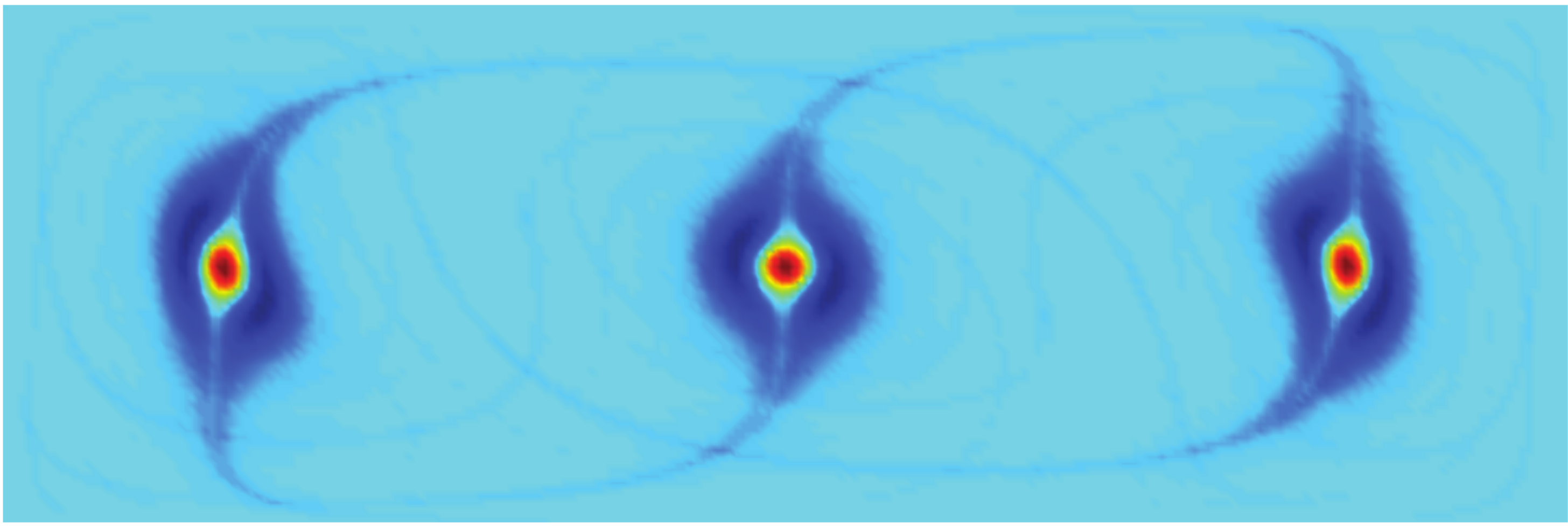}}
\subfigure[$v_5$]{
\includegraphics[width=3in,height=1in]{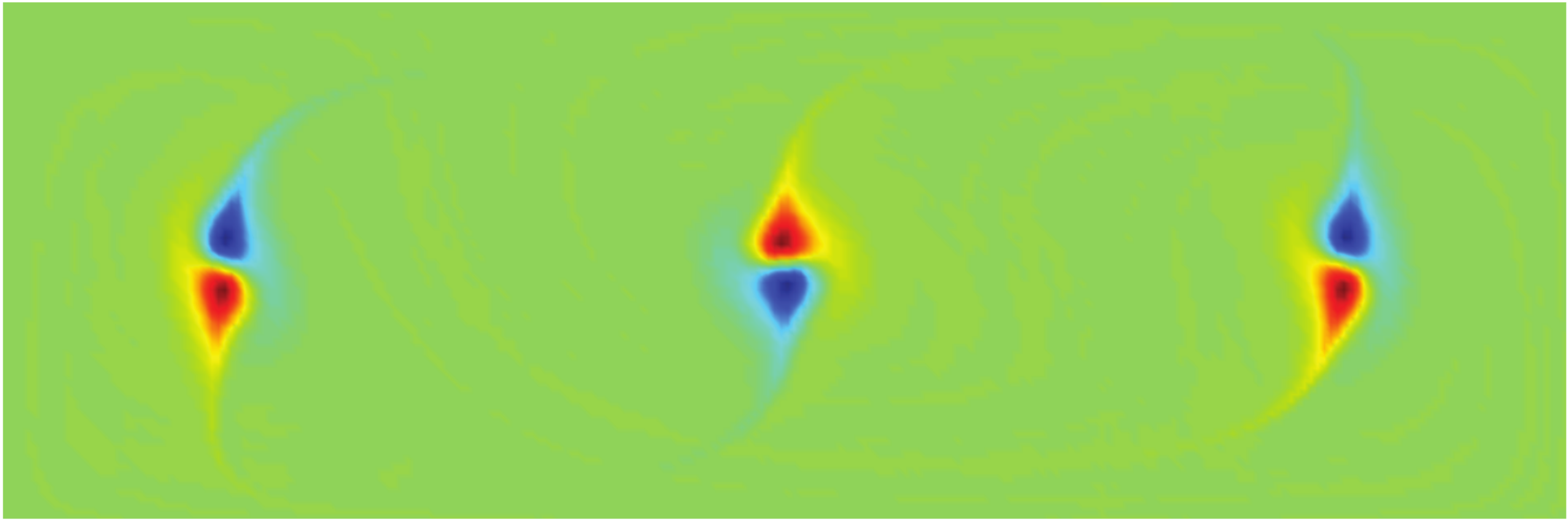}}
\subfigure[$v_6$]{
\includegraphics[width=3in, height=1in]{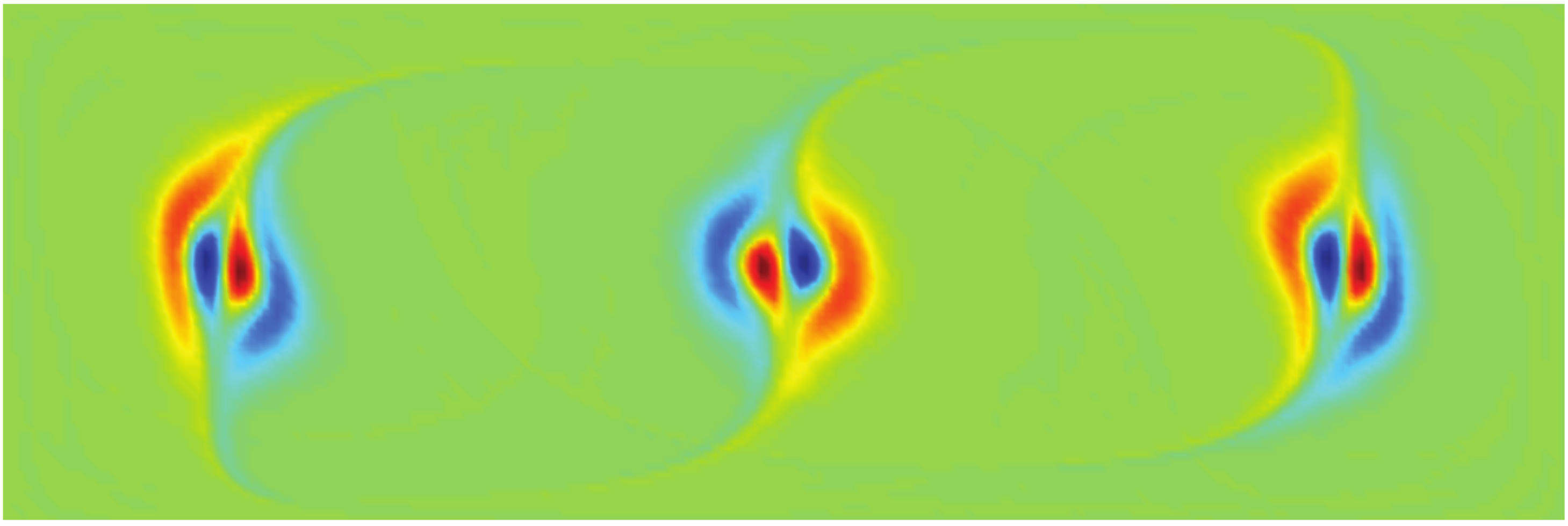}}
\end{center}
\caption{\footnotesize {The eigenvectors 3 through 6 of the reversibilized 
transfer operator, $R_{0,\tau_f}$, for $\tau_f\approx0.99$. Here red denotes highly positive regions, blue denotes highly negative regions, and green/yellow are close to 0.
}}
\label{fig:modes3to6}
\end{figure}

In figure \ref{fig:tau_cric}, we show the AISs based on $v_2$ for $\tau_f=0.99$. 
The collection of the three prominent regions corresponding to the set $A_1$ (i.e., the union of regions where $v_2>0$) forms an AIS. 
This almost invariant set  consists of three ACS components, which form a 3-stranded braid, as discussed next. 
The eigenvector that has $N$ ACS components, and hence forms an $N$-stranded braid,
is referred to as an $N$-stranded eigenvector in our discussion. Each (positive) lower eigenvalue has an associated eigenvector whose zero contour isolates other AIS, which are more `leaky', i.e., the lower the eigenvalue, the lower is the invariance of the associated AIS.
For reference, we show the next 4 eigenvectors in figure \ref{fig:modes3to6}, which tend to reveal smaller scale structures.



\subsection{Braiding of almost-cyclic sets}

Recall from Section \ref{sec:cases} that while the three period-3 fixed points cease to exist for $\tau_f < 1$ (see figure \ref{fig:poin_101}(d)), we can see from figure \ref{fig:tau_cric} that the three ACS, each of period-3, still exist in the same region in 
phase space for $\tau_f \approx 0.99$. 
In what follows, we examine the trajectories of these sets in space-time in order to determine if the corresponding braid is of PA type.  
 We use $\tau_f\approx0.975$ for our analysis in this case. 
To construct the physical braid for the ACS trajectories, we form a discretized family of time-shifted versions of the transfer operator 
$P_{t_i,\tau_f}$ for $t_i=(i-1) \tau_f/k$, where $i=1,2,\ldots,k$.  That is,
we are considering transfer operators corresponding to a family of period-$\tau_f$ flow maps, (\ref{flowmap}), i.e., 
\[
\phi^{t_i + \tau_f}_{t_i}: D \rightarrow D,
\] 
of different initial phases $t_i \in [0,\tau_f)$.
 
\begin{figure}[ht!]
\centering
\includegraphics[width=\textwidth]{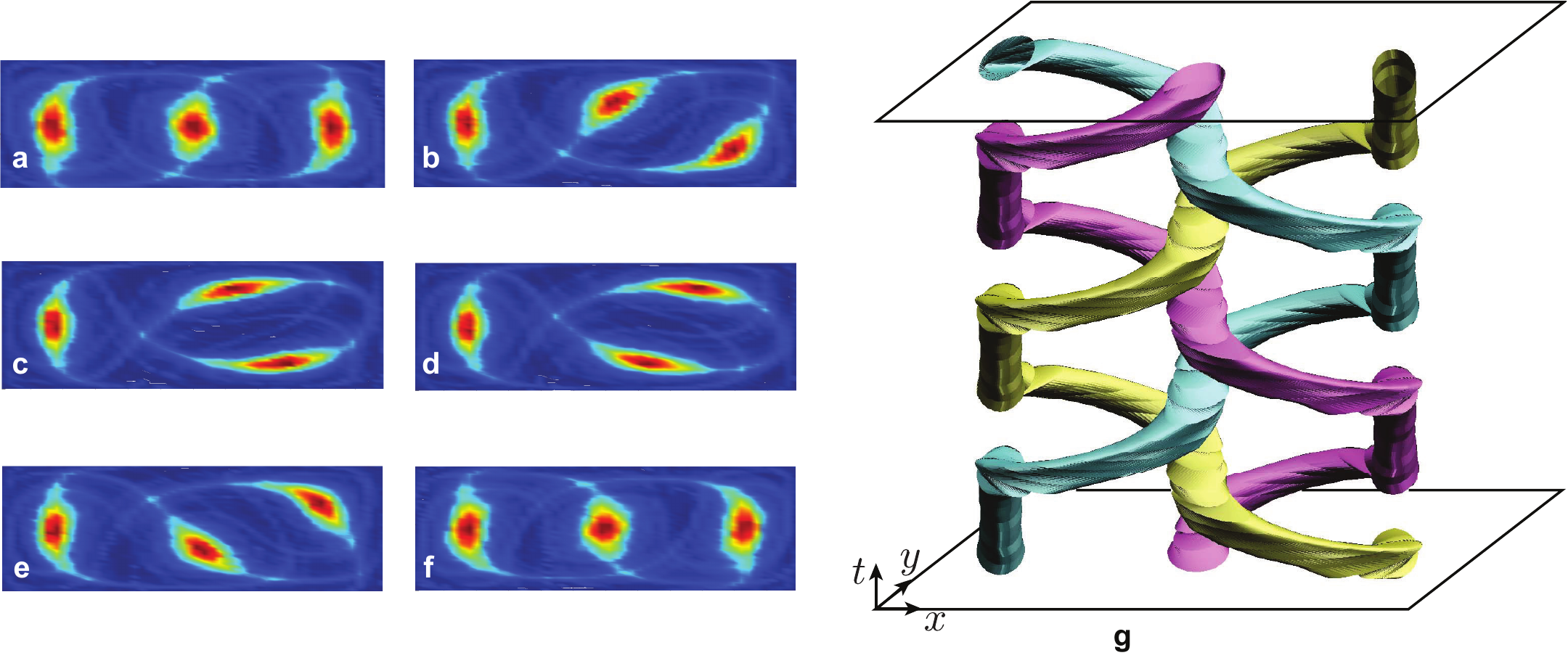}
\caption{\footnotesize{Braiding in (2+1)-dimensional space-time via three almost-cyclic sets for $\tau_f\approx 0.975$, a parameter value for which corresponding periodic points no longer exist (enhanced online). 
The 2nd eigenvector, $v_2$, of $R_{t,\tau_f}$ is shown for different phases for the first half-period, specifically $t/\tau_f=$ 
(a) $0$,
(b) $0.1$,
(c) $0.2$,
(d) $0.3$,
(e) $0.4$,
(f) $0.5$.
(g) The 3 ACSs shown braiding in space-time for 3 periods of flow (enhanced online). The braid is isotopic to the braid shown in figure \ref{cavity_flow_streamlines}(c).
}}
\label{fig:3dbraid}
\end{figure}

For each of the corresponding $R_{t_i,\tau_f}$, we find the almost-invariant structure based on the second eigenvector as mentioned previously, which reveals the three period-3 ACS for $\tau_f\approx0.975$. We take $k=20$, and in figure \ref{fig:3dbraid}(a-f) we show these structures for the first half-period of the flow.
This 
physical braid for this half-period of motion corresponds to the mathematical braid generator $\sigma_2$.
 Similarly, for the second half of the time-period of the flow (not shown), the stirring protocol is the one given by the braid generator $\sigma_{-1}$. Hence, the physical braid formed by the ACS trajectories for $\tau_f\approx0.975$ is represented by the mathematical braid $\sigma_{-1}\sigma_2$, which is identical to the mathematical braid formed by periodic points for $\tau_f=1$ or by elliptical islands for $\tau_f\geq1$. Three periods of the flow (one period of the braid) are shown in
 (2+1)-dimensional space-time 
in figure \ref{fig:3dbraid}(g), which can be seen to be isotopic to the braid shown in figure \ref{cavity_flow_streamlines}(c). Hence the lower bound on topological entropy is again given by $h_{\rm TN}\approx0.962$, and we can see from figure \ref{fig:entropy} that this lower bound is still valid.


\subsection{Persistence and bifurcation of almost-invariant sets} \label{br_bif}

\begin{figure}[h]
\includegraphics[width=\textwidth]{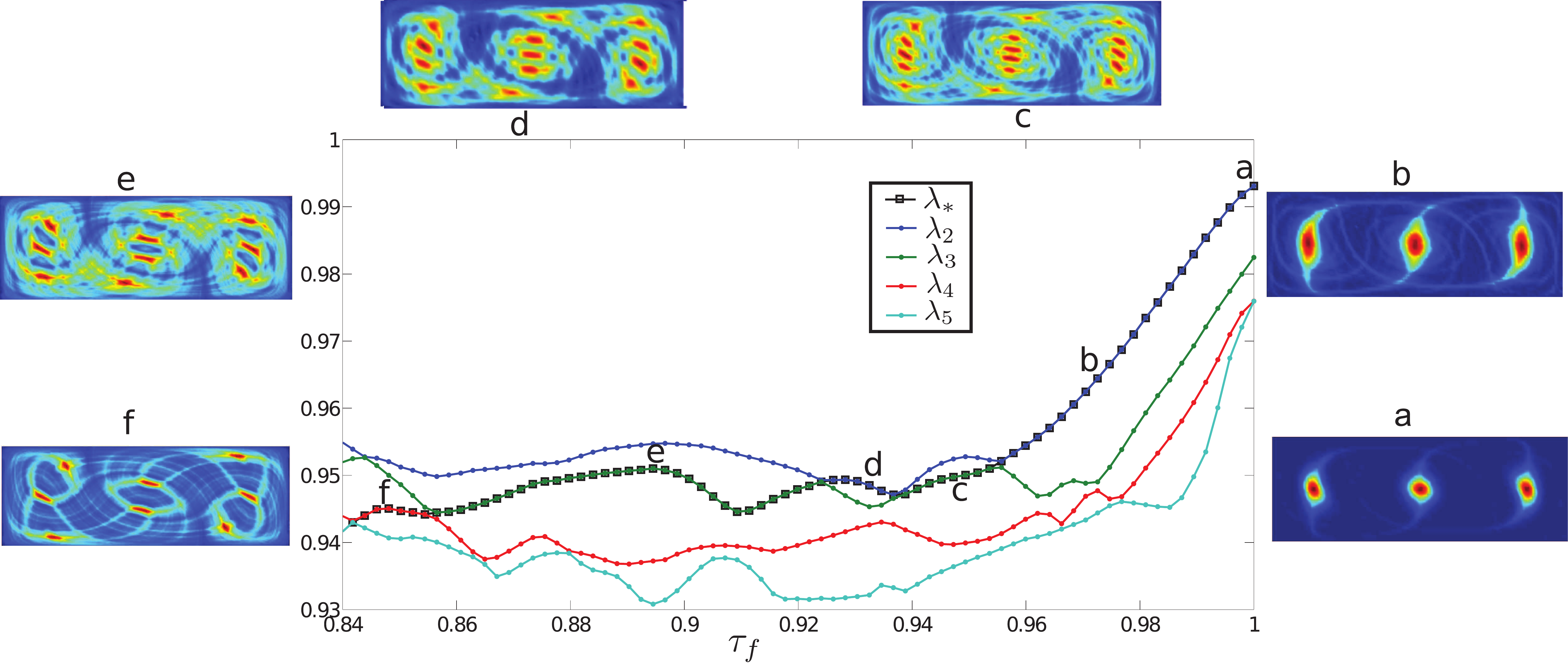}
\caption{\footnotesize Almost-invariant set structure via continuation of the 3-strand eigenvector found at (a) for various values of $\tau_f$. The structures shown in (a) and (b) can clearly be seen to again have 3 strands. The structure in (c) consists of 16 strands, while (d) can be seen to have 13 strands. (e) has 10 strands while (f) consists of 8 clearly identifiable strands. Also shown are the four largest eigenvalues of the reversibilized discretized transfer operator $R_{0,\tau_f}$ (except 1), $\lambda_2 > \cdots > \lambda_5$, which are colored according to their ordering.
One of these is singled out as the $\lambda_*$ branch (shown with large squares), which is the eigenvalue corresponding to $v_*^{\tau_f}$, the continuous eigenvector family of interest (see text for details).}
\label{fig:eigenvalue_bifurcation}
\end{figure}

As the parameter $\tau_f$ is varied, one expects to see continuous variation in the eigenvalues of  $R_{t,\tau_f}$.
Since the eigenvalues of $R_{t,\tau_f}$ are independent of the phase $t$, we can consider the initial phase case, $R_{0,\tau_f}$.
In figure \ref{fig:eigenvalue_bifurcation}, we plot the first few eigenvalues of $R_{0,\tau_f}$ in dependence on the parameter $\tau_f$, similar to  \cite{JuMaMe2004}.
Although several `crossings' of eigenvalue curves seem visible, care must be taken to determine if these are genuine crossings or simply close approaches, as eigenvalues generically `avoid' crossings if there is no symmetry present \cite{DeMe1994}.
Our concern is not necessarily to resolve families of eigenvalue curves, i.e., to determine whether genuine eigenvalue curve crossings occur.
Instead, we are interested in determining families of eigenvectors by the method of continuation.
In particular, we want to determine the family of eigenvectors that 
generate a consistent trend of PA braiding  
for $\tau_f\leq 1$.

We find that such a structure can be found by continuity of the branch of eigenvectors of $R_{0,\tau_f}$ that are connected to $v_2$ of $R_{0,1}$, 
since $v_2$ of $R_{0,1}$ captures the braiding structure correctly for the reference case. The continuation of this 3-strand eigenvector for $\tau_f<1$ is carried out as follows. 

First, we discretize the parameter space 
$0.8<\tau_f\leq 1$ 
into intervals of size $\Delta \tau_f   \approx 0.0021$. 
Then we calculate the reversibilized Perron-Frobenius operator $R_{0,\tau_f}$ for each of those parameter values. We compute the top 10 non-trivial eigenvectors (normalized to unity) 
$(v^{\tau_f}_2,..,v^{\tau_f}_{11})$ for each $\tau_f$. 
%

Our aim is to identify an eigenvector for each $\tau_f$ that can be used to obtain an AIS whose corresponding ACSs give a braiding structure similar to the reference case.
We denote this eigenvector by $v^{\tau_f}_*$. By definition, for $\tau_f=1$ the eigenvector that gives the relevant braiding structure is the second eigenvector, thus $v^{1}_*=v^{1}_2$. 
We calculate the inner product of $v^{1}_* $ with each of the 10 eigenvectors of the next lowest $\tau_f$ value, i.e., $\tau_f=1-\Delta \tau_f$, and we compute the absolute value of the inner product, $\beta_j = | \langle v^1_*,v^{1-\Delta \tau_f}_j\rangle|$, for $j=2,...,11$.  
We select the eigenvector $v^{1-\Delta \tau_f}_j$ 
that gives the highest value of $\beta_j$, 
and refer to this eigenvector as $v^{1-\Delta \tau_f}_*$. 
We denote this value as $\beta_{\tau_f}$, i.e., $\beta_{\tau_f}=|\langle v^{\tau_f}_*,v^{\tau_f-\Delta\tau_f}_*\rangle|$.
Similarly, this procedure is carried out for $\tau_f=1-2\Delta\tau_f$ by taking the norm of the inner products with $v^{1-\Delta\tau_f}_*$, and so on. 

Using this procedure, we make some observations based on figure \ref{fig:eigenvalue_bifurcation}. The AIS structure consisting of 3 ACSs seems to persist until $\tau_f\approx 0.96$. We also observe that as the value of $\tau_f$ decreases further, the structure consisting of 3 ACSs starts breaking up, and at $\tau_f\approx0.95$ a structure consisting of 16 ACSs can be seen, as shown in figure \ref{fig:eigenvalue_bifurcation}(c). Further decreasing $\tf$ produces another change in the AIS structure, and figure \ref{fig:eigenvalue_bifurcation}(d) shows an AIS with 13 ACSs. Moving further to the left on the $\tau_f$ axis, we see AIS structures with 10 and 8 ACSs in figures  \ref{fig:eigenvalue_bifurcation}(e) and \ref{fig:eigenvalue_bifurcation}(f), respectively.

\begin{figure}[t!]
\begin{center}
\includegraphics[width=0.75\textwidth]{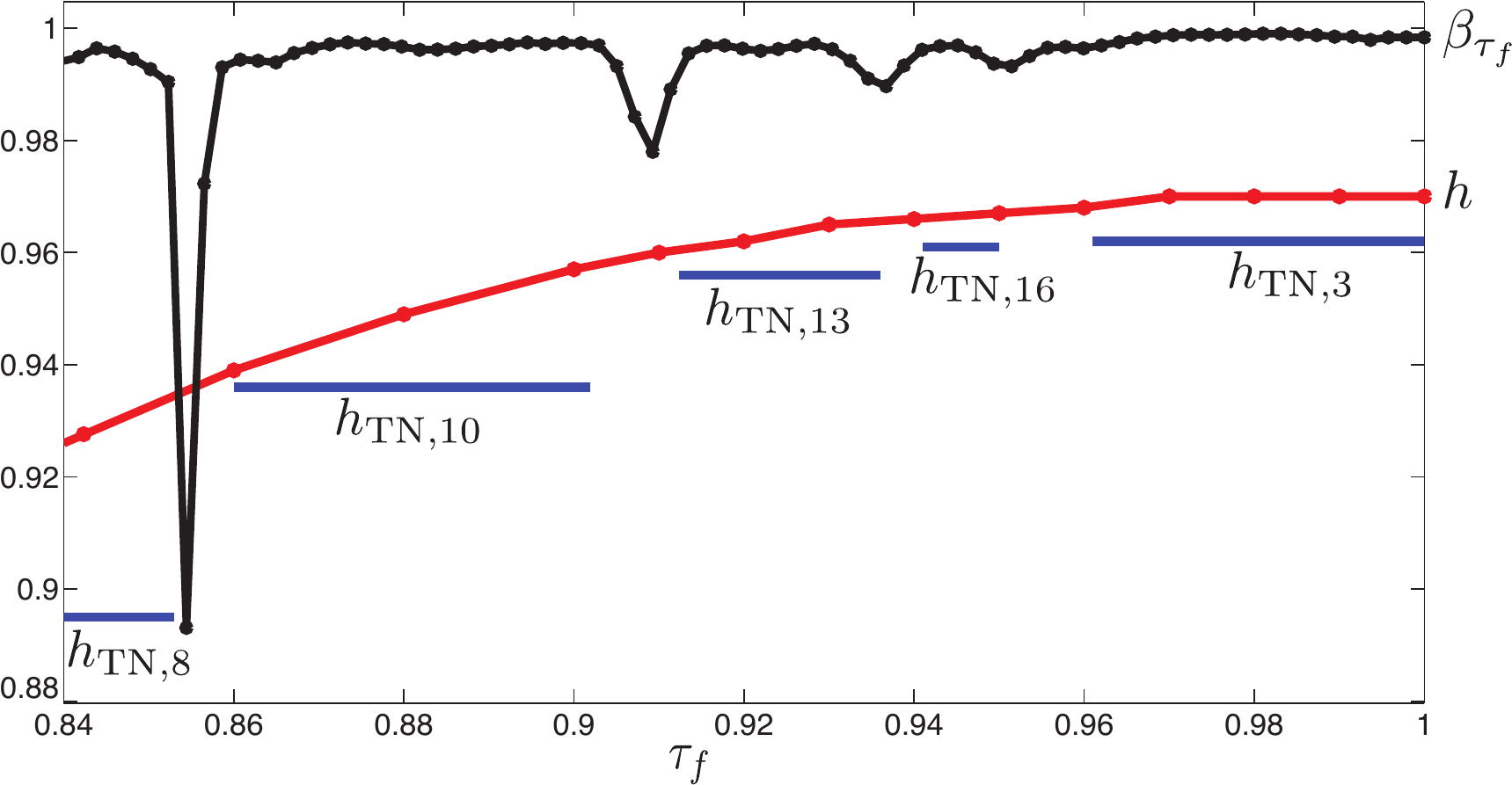}
\end{center}
\caption{\footnotesize 
$\beta_{\tau_f}$ for a range of $\tau_f$ values (black curve). The regions where this value stays almost constant are the regions of persistence of different braiding structures, separated by transition regions where the value drops significantly. The topological entropy of the flow, $h_f$, computed by line stretching, reproduced from figure \ref{fig:entropy}, is shown in red. The lower bounds computed by the TNCT for the braids generated by the ACSs in that parameter range are shown in blue in the relevant regions.}
\label{fig:dotprod}
\end{figure}

This break-up of the AIS structure can be better understood by plotting the $\beta_{\tau_f}$ values, shown in figure \ref{fig:dotprod}. 
The $\beta_{\tau_f}$ value measures the closeness of the eigenvectors along the $v_*$ branch for two nearby $\tau_f$ values, and hence captures changes in the eigenvector morphology. We observe that in regions where we can clearly identify the different AIS structures consisting of 3, 16, 13, 10 or 8 ACSs, the value of $\beta_{\tau_f}$ remains almost constant. On the other hand, in the transition regions where there is no clearly distinguishable AIS structure there is a sharp drop in the $\beta_{\tau_f}$ value, signifying a significant change in the structure of the eigenvector. Intuitively, this behavior of $\beta_{\tau_f}$ means that 
the AIS structure persists for a range of $\tau_f$ and then undergoes a transition (or bifurcation) into another AIS structure. Comparing figure \ref{fig:eigenvalue_bifurcation} with figure \ref{fig:dotprod}, we see that the dips in the graph of $\beta_{\tau_f}$ versus $\tau_f$ seem to accurately capture the boundaries of different AIS structures and transition regions.
The ACS found using the above method can each be identified with a braid, with the number of strands in the braid corresponding to the number of ACS in the domain.  For parameter ranges over which  $\beta_{\tau_f}$  remains essentially constant, we find that the number of strands in each braid remains constant and the braids are isotopic to each other.  We also find that  dips in the value of $\beta_{\tau_f}$  correspond to changes in the number of strands in the representative braid and changes in the braid topology.  
We determine the space-time braid structure produced by the ACSs in each case by computing the time-shifted transfer operators as was done for the braid on 3 strands in the previous subsection.  

\begin{figure}[h]
\includegraphics[width=\textwidth,height=0.25\textwidth]{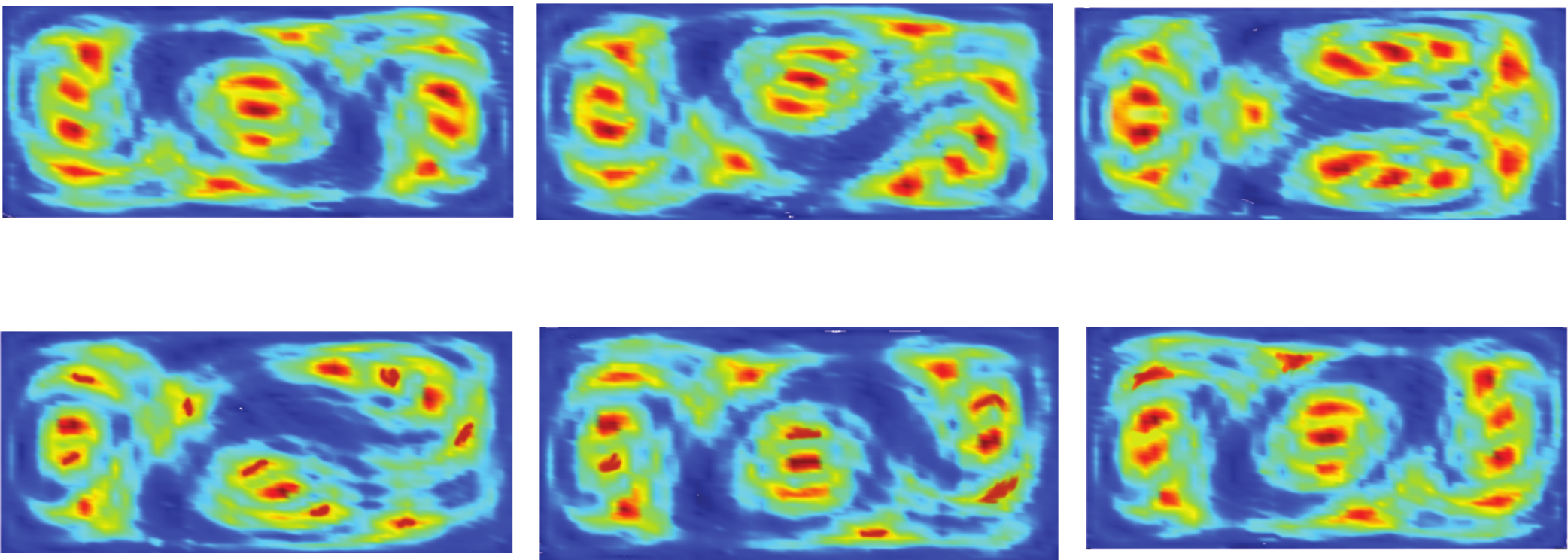}
\caption{\footnotesize{AIS identified by second eigenvectors of time-shifted $R_{t_i,\tau_f}$ for the first half of the flow period $\tau_f$, where $\tau_f \approx 0.93$ (enhanced online). Left to right:  Top-row: $t_i/\tau_f$= $0$, $0.1$, $0.2$. Bottom row: $t_i/\tau_f$= $0.3$, $0.4$, $0.5$. The braiding motion of the 13 ACS  is  evident, and is illustrated in figure~\ref{fig:3dbraid}.}}
\label{fig:13braid_evecs}
\end{figure}

\begin{figure}[ht!]
\begin{center}
\subfigure[Initial state]{
\includegraphics[width=2.5in]{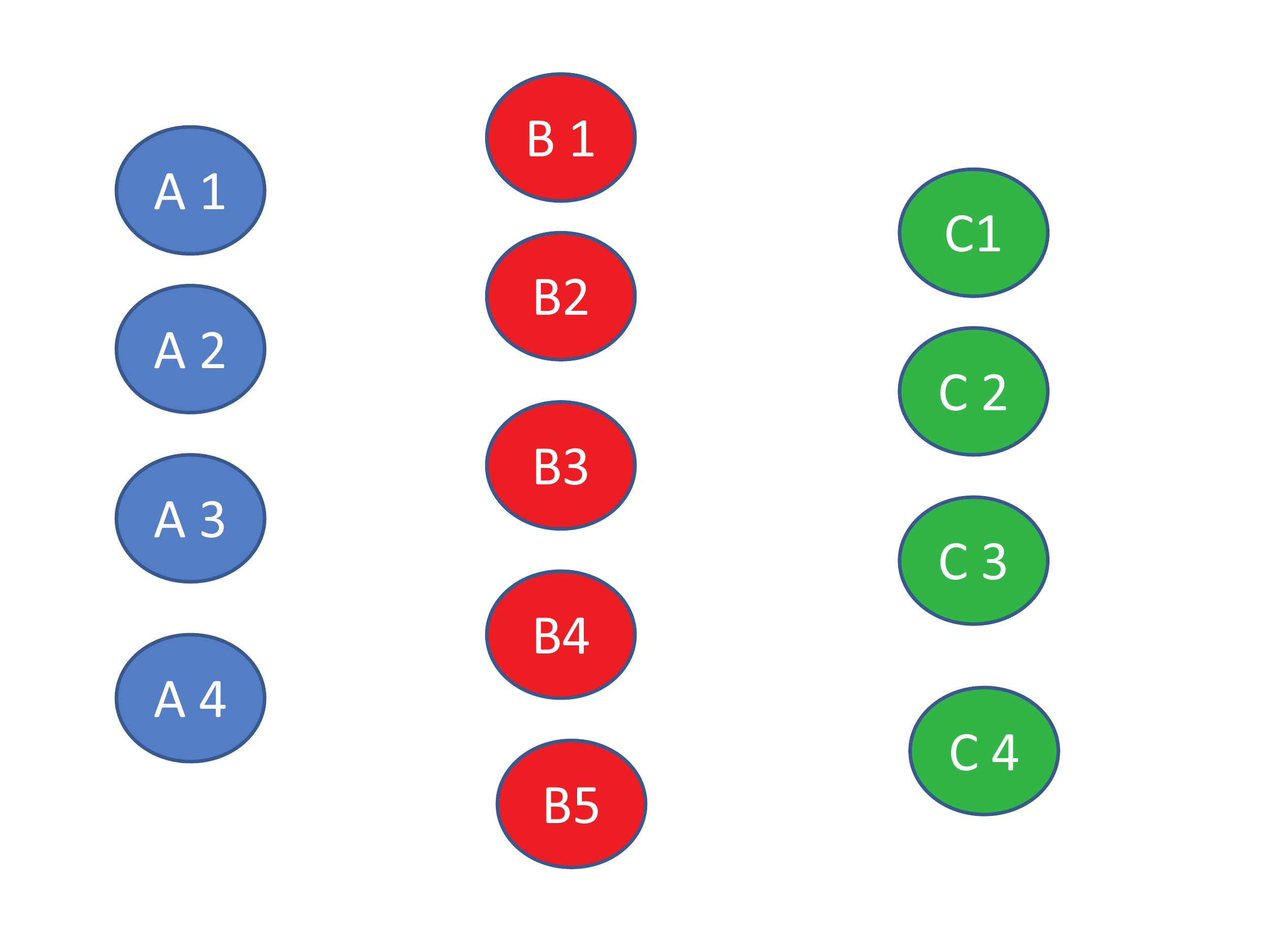}}
\subfigure[Movement during pulse 1]{
\includegraphics[width=2.5in]{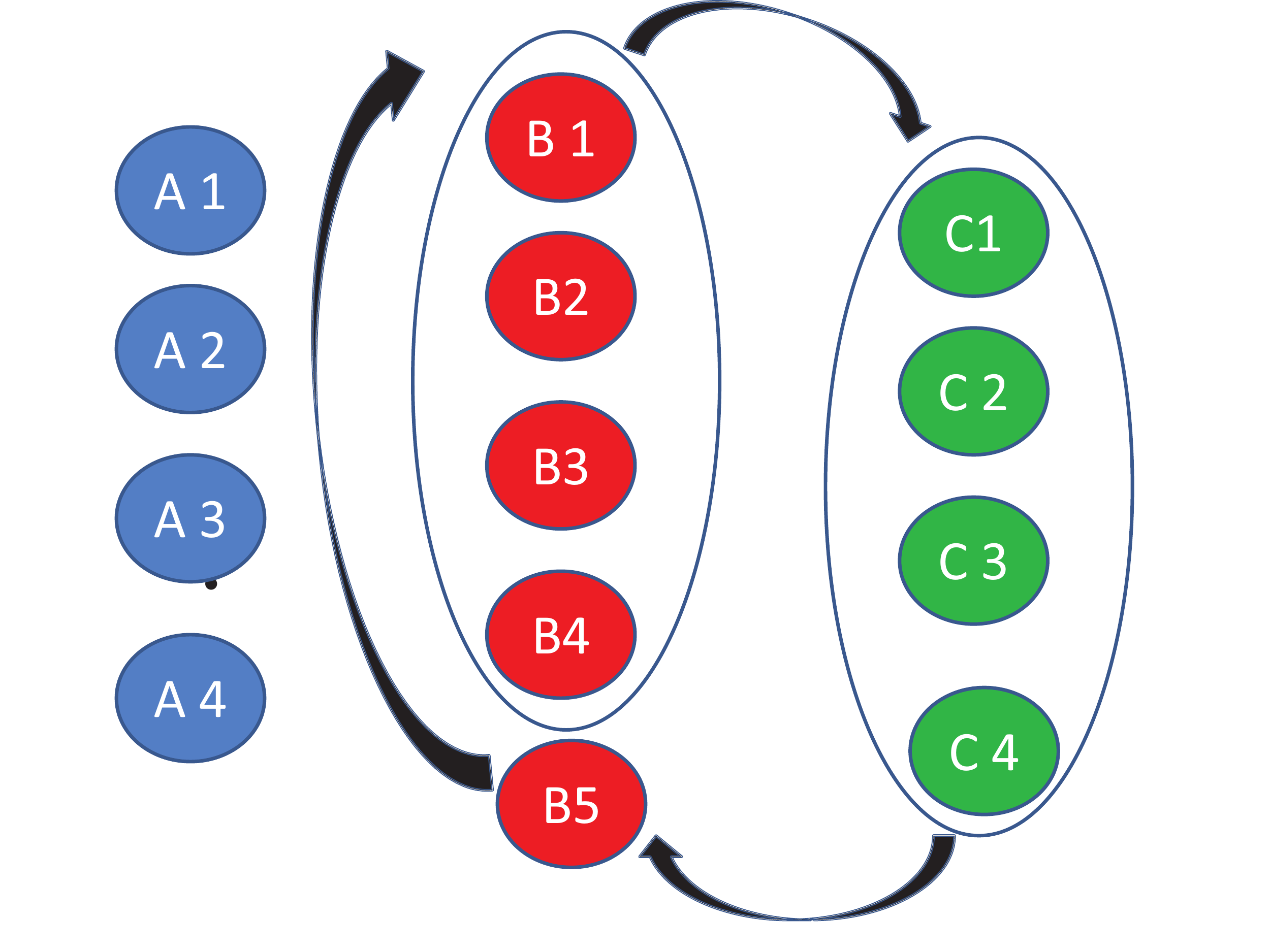}}\\
\subfigure[Movement during pulse 2]{
\includegraphics[width=2.5in]{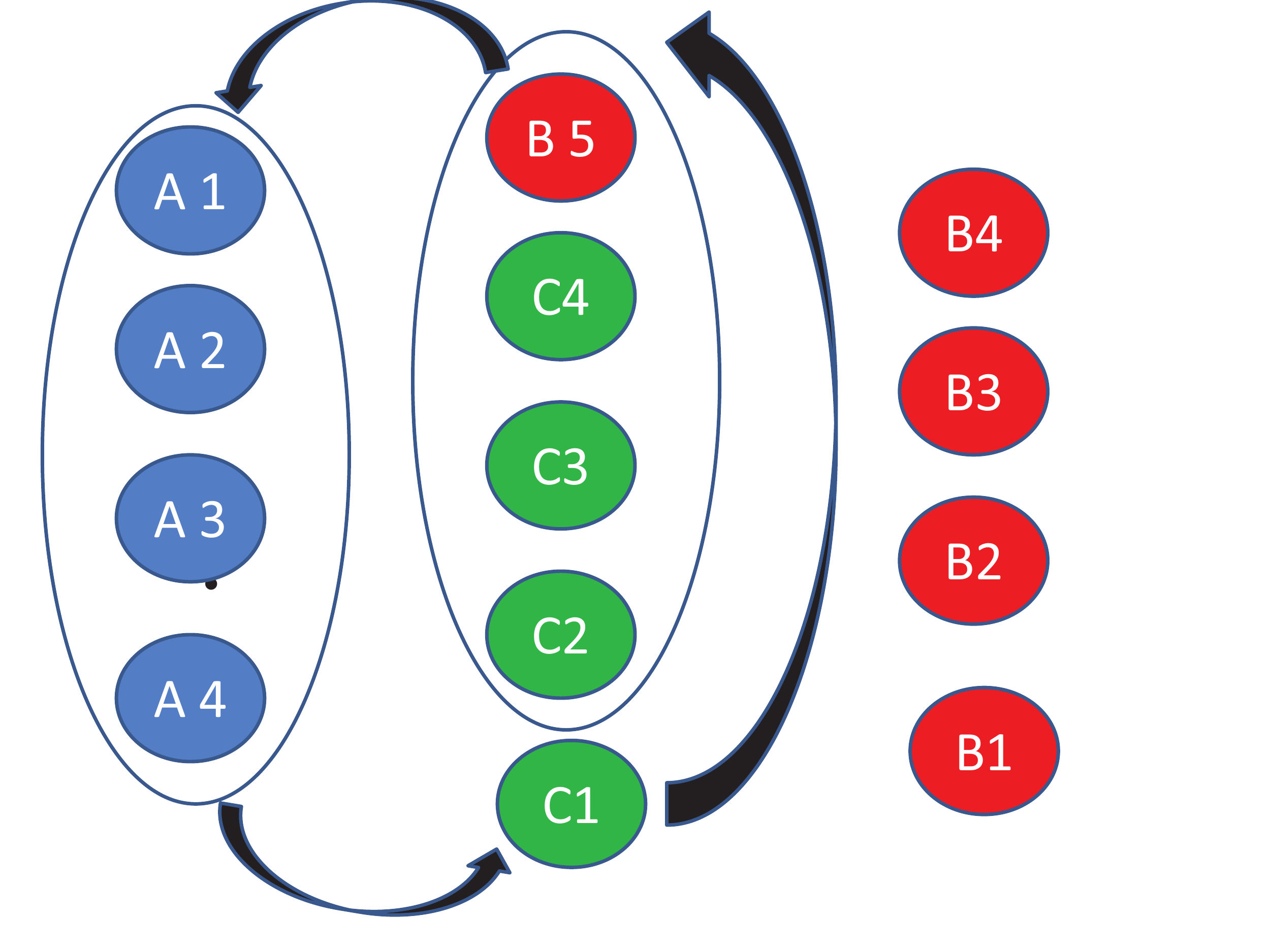}}
\subfigure[State after 1 period (i.e., 2 pulses)]{
\includegraphics[width=2.5in]{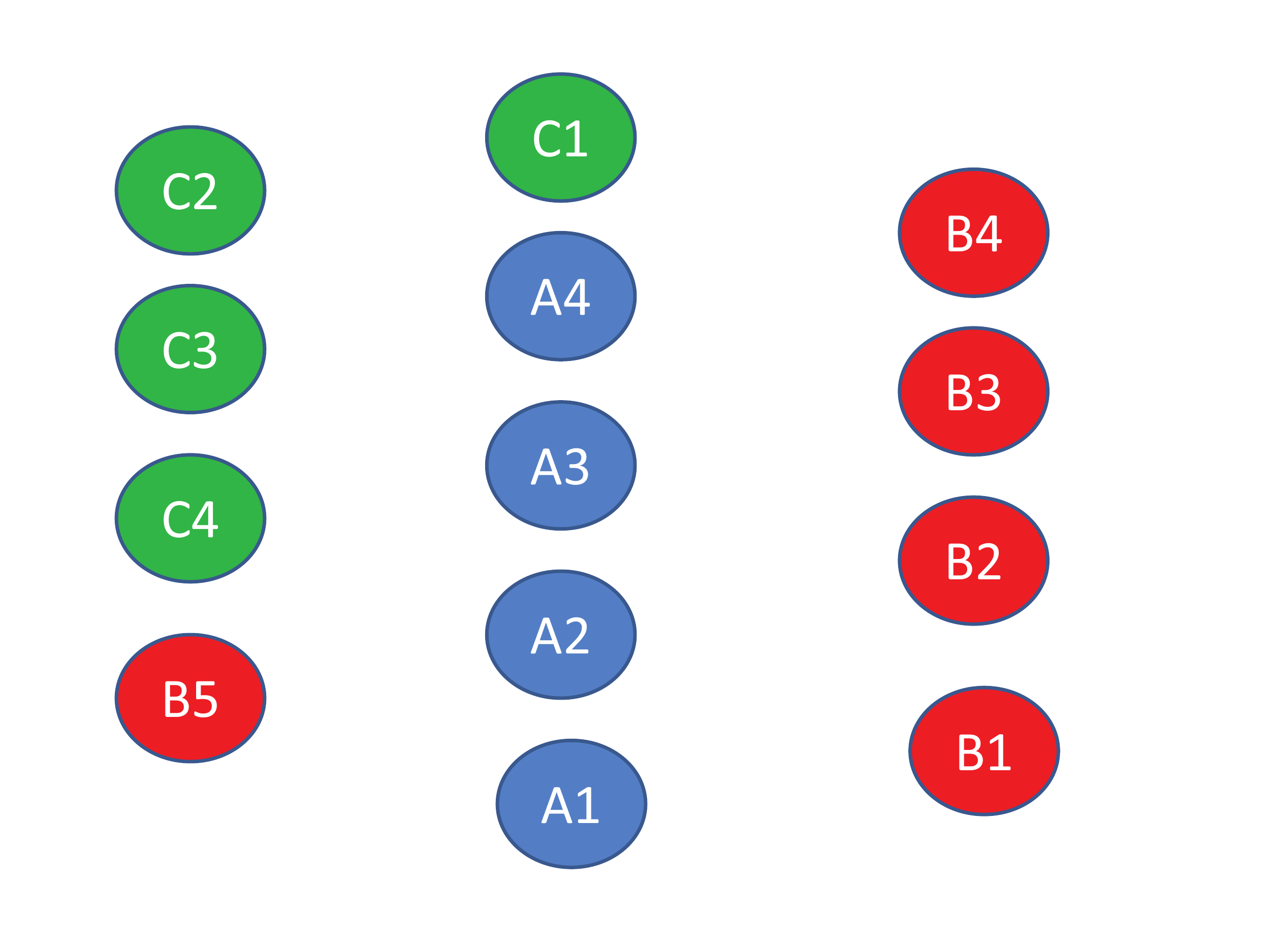}}
\end{center}
\caption{\footnotesize {Movement of the 13 ACSs identified from figure~\ref{fig:13braid_evecs} over one period of the flow.  The ACSs return to their initial positions after 13 periods. Note that what was formerly the central set (strand) in the reference case now consists of five disjoint sets (strands), shown in red. The strand B5 is `left behind' in the center of the domain during the first half-period and subsequently moves to the left during the second half-period, generating a topologically distinct braid from that shown in  figure \ref{fig:3dbraid}.  We find this braiding motion of the ACS to persists for roughly $0.910\leq\tau_f\leq 0.935$. 
}}
\label{13braid_illus}
\end{figure}

\begin{figure}[h!t]
\centering
\subfigure[]{\includegraphics[height=4.75in, width=3in]{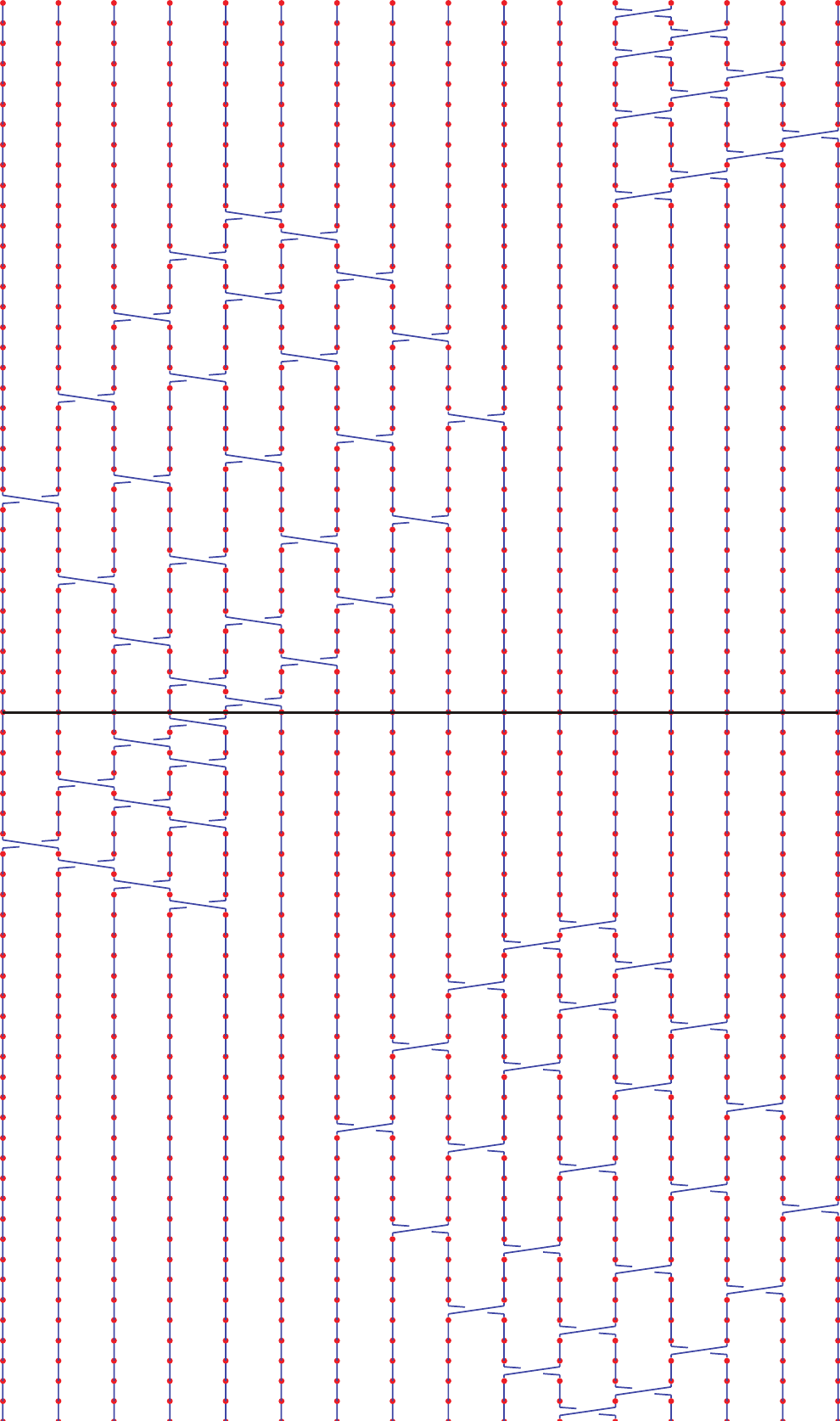}}
\hspace{.25in}
\subfigure[]{\includegraphics[height=4.75in,width=3in]{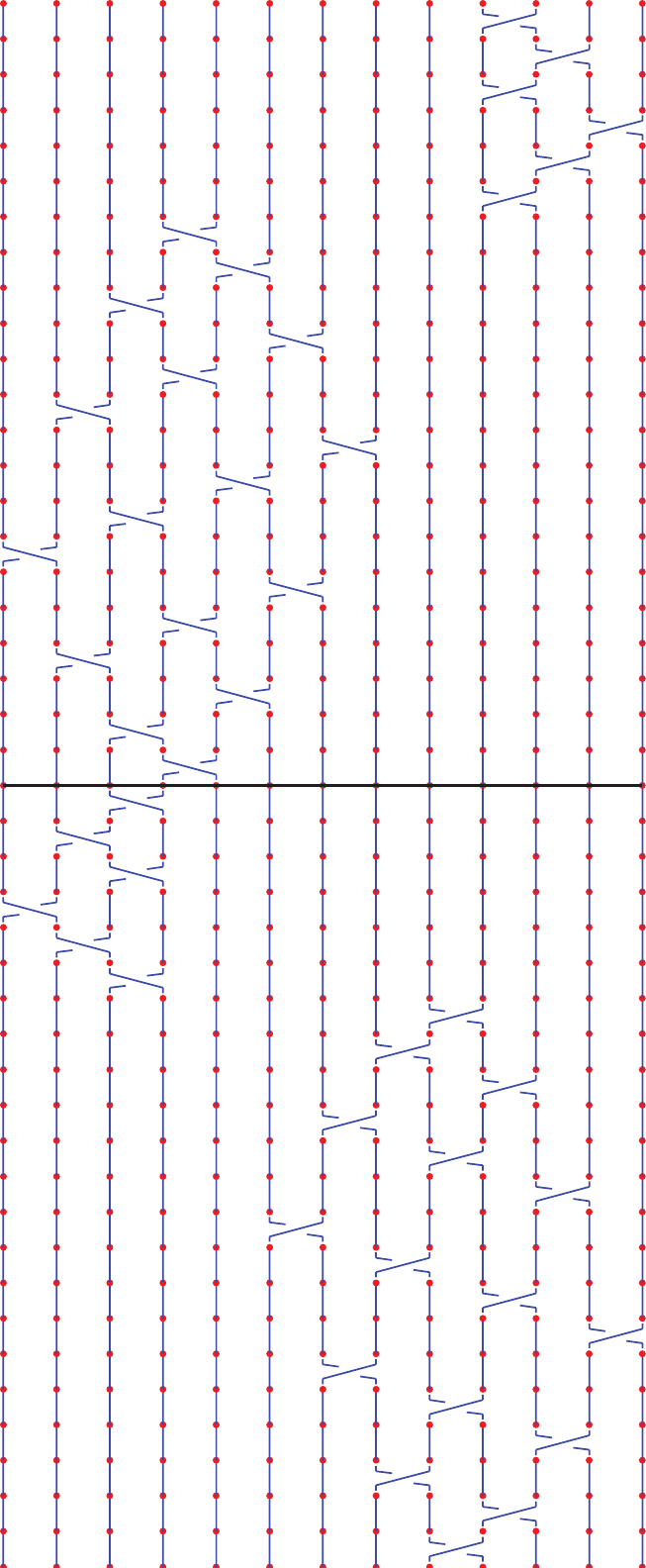}}
\subfigure[]{
\includegraphics[height=2.25in,width=2.75in]{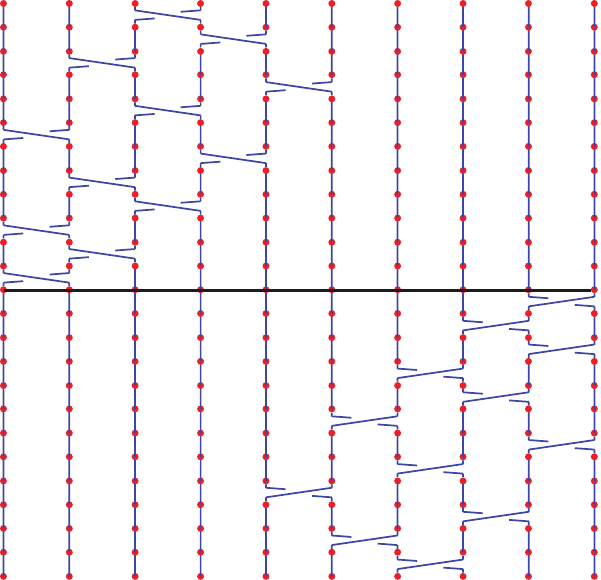}}
\hspace{.5in}
\subfigure[]{
\includegraphics[height=2.25in,width=2.75in]{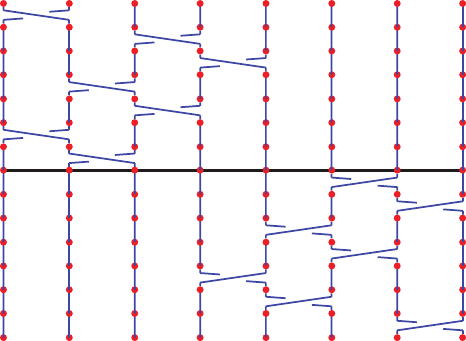}}
\caption{\footnotesize {Physical representations of different braids. The time is increasing from bottom to top. (a) The braid on 16 strands, $0.941 \le \tf \le 0.949$. 
(b) The braid on 13 strands, $0.910\leq\tau_f\leq 0.935$. 
(c) The braid on 10 strands, $0.861\leq\tau_f\leq 0.890$. 
(d) The braid on 8 strands,  $0.840\leq\tau_f\leq 0.852$. 
}}
\label{fig:braid_rep}
\end{figure}

Consider the case $\tau_f \approx 0.93$, for which the ACSs form a 13-stranded braid.  
In figure~\ref{fig:13braid_evecs}, we show the time-shifted eigenvectors at several different times during the first half-period of the flow, 
and we illustrate the motion of the 13 strands during one full period of the flow in figure~\ref{13braid_illus}. 
The overall structure of the braid consists of four strands each on the left and right of the domain, and five strands in the middle. During the first half-period, strands B1 through B4 move together to the right while C1 through C4 move to the left, similar to the $\s{2}$ motion of the middle and right strands in the   reference case, but strand B5 remains in the middle. Similarly, during the second half-period,  strand C1 remains in the middle, while C2 through C4 and B5 move to the left. 
This motion of the ACS gives rise to the physical braid representation shown in figure \ref{fig:braid_rep}(a), which can be written as $\sigma_L\sigma_R$, where
\begin{subequations}
\begin{equation}
\small
	\sigma_R = \sigma_{-3}\, \sigma_{-2}\, \sigma_{-3}\, \sigma_{-1}\, \sigma_{-2}\, \sigma_{-3}\, \sigma_9\, \sigma_8\, \sigma_{10}\, \sigma_7\, \sigma_9\, \sigma_{11}\, \sigma_6\, \sigma_8\, \sigma_{10}\, \sigma_{12}\, \sigma_7\, \sigma_9\, \sigma_{11}\, \sigma_8\, \sigma_{10}\, \sigma_9
\end{equation}
is the braid over the first half period of the flow, and
\begin{equation}
\small
	\sigma_L = \s{10}\, \s{11}\, \s{10}\, \s{12}\, \s{11}\, \s{10}\, \s{-4}\, \s{-5}\, \s{-3}\, \s{-6}\
	 \s{-4}\, \s{-2}\, \s{-7}\, \s{-5}\, \s{-3}\, \s{-1}\, \s{-6}\, \s{-4}\, \s{-2}\, \s{-5}\, \s{-3}\, \s{-4}
\end{equation}
\end{subequations}
is the braid over  the second half period.  
All strands return to their original positions after 13 periods of the flow. This motion is hence topologically distinct from the 3-stranded reference braid shown in figures~\ref{cavity_flow_streamlines}(c) and \ref{fig:3dbraid}(g). 
The lower bound on the topological entropy of this braid is  $h_{{\rm TN},13}\approx0.956$, which is lower than $h_{{\rm TN},3}$ for the reference case.

Over the range of parameters considered here, $0.84\le\tf\le 1.0$, we find that decreasing $\tf$ from unity gives ACSs that generate braids on 3, 16, 13, 10, and 8 strands, successively.  The physical braid representations are shown in figure \ref{fig:braid_rep}, and the corresponding  braid words  are given in the appendix.   The basic behavior in all of the bifurcations tends to follow the pattern discussed in the 13-strand case --- one of the strands from the middle of the domain breaks away from the rest of the strands, and hence we get topologically different braiding in each case. 
Table \ref{entropy_table} lists the topological entropy of these braids, and these values are plotted in figure~\ref{fig:dotprod}. 
%
%
\begin{table}
\begin{center}
    \begin{tabular}{ |c|c|c|}
    \hline
	    $\tf$ range & $N$ & $h_{{\rm TN},N}$ \\ \hline
	    $0.960 \le \tf \le 1$ 		& 3 	& 0.962\\ \hline
	    $0.941 \le \tf \le 0.949$ 	& 16 & 0.961\\ \hline
	    $0.910\leq\tau_f\leq 0.935$	& 13 & 0.956\\ \hline
	    $0.861\leq\tau_f\leq 0.890$	& 10	& 0.936\\ \hline
	    $0.840\leq\tau_f\leq 0.852$	& 8 	& 0.894\\ \hline
    \end{tabular}
    \caption{\footnotesize{
    Topological entropy values, $h_{{\rm TN},N}$, for the braids on $N$ strands generated by the ACSs for the various $\tf$ values considered here.}}\label{entropy_table}
 \end{center}
 \end{table}
We see that the lower bounds provided by the TNCT for these braids, ${\htn}_{,N}$, are indeed sharp lower bounds on the actual topological entropy for the flow, $\hf$.  Furthermore, ${\htn}_{,N}$ is a good estimate of $\hf$ in these cases.
This is a key result and suggests 
that ACS-based topological analysis can be used in the framework of the TNCT. 

\section{Conclusions and Future Directions}

We give numerical evidence that the almost-invariant set theory can be applied in conjunction with the Thurston-Nielsen classification theorem to find estimates of complexity in a wide class of systems. 
We have used 
almost-invariant sets as central objects in the 
application of the TNCT 
by looking at  non-trivial braids  corresponding to the space-time trajectories of these sets.
We demonstrate that this procedure can predict an accurate lower bound on the topological entropy for a flow even in those cases when the system has been perturbed far from one that contains braiding periodic orbits.  
This work shows that predictions regarding chaos can be made by considering regions in phase space that move together for a finite amount of time. 
This application broadens the scope of problems where such a lower bound can be established and suggests that they should be of use in time varying systems that do not have any well-defined periodic points, such as free surface flows, or in time-independent systems  operating in a parameter regime where there do not exist any easily-identifiable fixed points.

We stress the fact that while the TNCT can be (in theory) applied to any orientation-preserving flow (on a two-dimensional domain) that results in a diffeomorphism, the difficulty lies in finding the appropriate isotopy class. In the cases where low-order periodic orbits exist (and can be found), the isotopy class is defined relative to the periodic orbits. In these cases, the stretching produced by the TN representative is a strict lower bound on the stretching for any flow in the isotopy class.  For the cases in which no obvious, low-order periodic orbits exist and the braiding is instead identified via almost-cyclic sets, the predictions of stretching are only approximate, since the ACSs do not rigorously qualify as punctures in phase space. However, this ACS approach 
identifies a small set of `approximate punctures' that 
give an accurate representation of the topological entropy for the cases we have considered.

%

%
 
The class of set-oriented transfer operator methods discussed in this paper have been applied to systems accessible solely by time series data \cite{FrPaEnTr07}, systems with stochasticity \cite{billings2008c}, and non-autonomous systems defined by data over finite time, where finite-time coherent sets become the objects of interest \cite{FrSaMo2010}. Hence, the connection between set-oriented statistical methods and topological methods in dynamical systems made in this paper provides an additional tool for analyzing complex systems, including those defined by data.
Ideally, one would want to apply the notion of braiding coherent sets to an experimental setting, where one could ask questions regarding the role of braiding coherent sets in complex physical systems.  For example, what implications does large-scale braiding have for stirring in geophysical flows? 

There is work on spectral analysis of the ``mixing matrix'' \cite{singh2009pf} that is closely related to the identification of almost invariant sets and ACSs, 
and connections have been made between this mixing matrix analysis  and the ``strange eigenmode'' that arises from spectral analysis of the continuous advection-diffusion operator \cite{LiuHaller2004,Froyland3}.  The topological information available from analyzing trajectories of ACS    suggests that a similar approach can be applied to the braiding of eigenvectors in these related methods.

\begin{figure}[!h]
\centering
\includegraphics[height=0.8\textwidth]{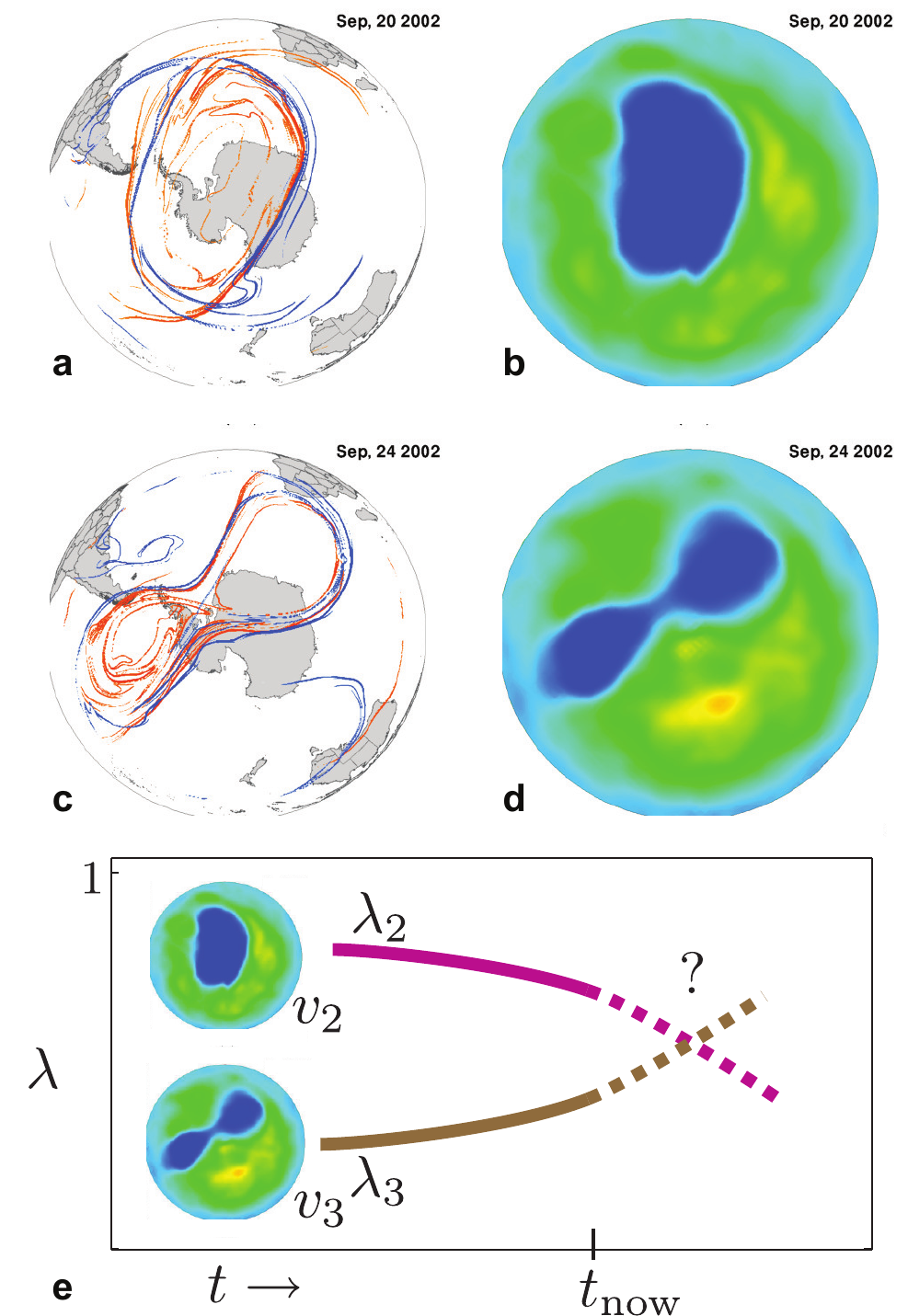}
\caption{\footnotesize {
(a,c) Atmospheric Lagrangian coherent set boundaries (LCSBs) 
for days surrounding an Antarctic polar vortex splitting event in September 2002 (based on NCEP/NCAR reanalysis data);
attracting (repelling) curves 
are shown in blue (red). Before and after the splitting event in late September, we see an isolated blob of air, bounded by LCSB curves, slowly rotating over Antarctica. 
The vortex pinches off, sending the northwestern part of the ozone hole off into the midlatitudes while the southwestern portion goes back to its regular position over Antarctica. 
We note that the  time-varying structures that bound the ozone hole fragments provide a framework for understanding the geometry of atmospheric transport; they act as analogs of stable and unstable manifolds of saddle-like almost cyclic sets.
(b,d) The corresponding daily ozone concentration (based on NASA TOMS satellite data).
(Adapted from \cite{LeRo2010})
(e) Shown in schematic is a possible change in ordering of the leading eigenvectors for the polar vortex, suggesting a new means to determine global bifurcations of a system. Even when the `single-blob' eigenvector is dominant, there may be `double-blob' eigenvector whose eigenvalue is increasing. Somewhere near the crossover point, the `double-blob'  eigenvector will be dominant and determine the large-scale characteristics of the flow.
	}}
\label{PolarVortex}
\end{figure}

We note that for the lid-driven cavity system, some dynamical `memory' of the saddle point stable and unstable manifolds remains even where {\it there are no more saddle points}; compare figure \ref{fig:lobe_dynamics}, where such objects exist, with the wisp-like features of figures \ref{fig:tau_cric} and \ref{fig:modes3to6}, where they do not technically exist.
Transfer operator methods seem to reveal structures resembling stable and unstable manifolds of invariant sets, but these are now associated with almost-invariant sets.
Given the importance of stable and unstable manifolds in applications, future work could extend this concept of stable and unstable `almost-invariant manifolds' to systems that are not exactly periodic, and possibly far from periodic. 
For example, in applications to stability and control \cite{SeKoLoMaPeRoWi2002, KoLoMaRo2008, MaRo2006, GrRo2009}, 
`almost-invariant manifolds' may prove useful for geometric interpretations of stable and unstable directions in the phase space.
They may also help in making connections with invariant manifold-like objects detected by other methods; cf.\ figure \ref{PolarVortex}.

Another intriguing possibility is opened up by considering the spectral dependence of transfer operators.
Even in relatively simple systems such as shown in figure \ref{fig:eigenvalue_bifurcation}, one sees interesting changes in the eigenvectors of the discretized Perron-Frobenius operator as a system parameter is varied. 
Some modes increase in importance while other modes decrease.
Different modes can correspond to dramatically different behavior.
This observation may have an interesting application to systems defined from data:\ prediction of dramatic changes in system behavior based on mode variations.
That is, a system may contain hints of a bifurcation before a bifurcation occurs, and this premonition may be teased out via a transfer operator approach, similar to  \cite{JuMaMe2004}.
To take a vivid example, consider figure \ref{PolarVortex}, where we show the splitting of the ozone hole in 2002 \cite{LeRo2010}, which we might  classify as a `Duffing bifurcation', where the bifurcation parameter is some physical quantity changing slowly with time.  
Given only the weather observations up to some time, $t_{\rm now}$, before the split, e.g., September 20, 2002 (figure
\ref{PolarVortex}(a,b)), might  the spectrum of the discretized Perron-Frobenius operator reveal an emerging, and topologically distinct, eigenvector which is increasing in importance, soon to become the dominant mode
(figure \ref{PolarVortex}(e))?
To determine trends in the change of spectrum of a transfer operator in real-time, one may need to consider the possibility of using the infinitesimal generator of the discretized Perron-Frobenius operator, which could give several orders of magnitude increase in speed for determining the spectral characteristics \cite{Froyland2011}. 


\section{Acknowledgments}
This work is dedicated to the memory of Jerry Marsden and Hassan Aref, mentors and friends who started us on the path that has led to this paper. We thank Mohsen Gheisarieha for helpful discussions, insights, and comments. We thank Phil Boyland for helpful discussions and providing a historical perspective of the field. We also thank the reviewers for the valuable comments that improved the quality of this paper. This material is based upon work supported by the National Science Foundation under Grant No.\ 1150456 to SDR. 



\begin{appendix}

\section{Braid words for different braids}

%

We give braid words for two pulses (i.e., one complete flow period) in each case. Here, $\sigma_R$ refers to first half of the time period and $\sigma_L$ refers to second half. The complete braid word for each braid is formed by composing the two words, i.e., $\sigma_L\sigma_R$.
\begin{itemize}
\item{Braid on 3 strands.\\ $\sigma_L=\sigma_{-1}$ \\ $\sigma_R=\sigma_{2}$}
\item{Braid on 16 strands.\\$\sigma_L=\sigma_{12}\sigma_{13}\sigma_{12}\sigma_{14}\sigma_{13}\sigma_{12}\sigma_{15}\sigma_{14}\sigma_{13}\sigma_{12}\sigma_{-5}\sigma_{-6}\sigma_{-4}\sigma_{-7}\sigma_{-5}\sigma_{-3}\sigma_{-8}\sigma_{-6}\sigma_{-4}\sigma_{-2}\\\
\hspace*{.45in}\sigma_{-9}\sigma_{-7}\sigma_{-5}\sigma_{-3}\sigma_{-1}\sigma_{-8}\sigma_{-6}\sigma_{-4}\sigma_{-2}\sigma_{-7}\sigma_{-5}\sigma_{-3}\sigma_{-6}\sigma_{-4}\sigma_{-5}$ \\$\sigma_R=\sigma_{-4}\sigma_{-3}\sigma_{-4}\sigma_{-2}\sigma_{-3}\sigma_{-4}\sigma_{-1}\sigma_{-2}\sigma_{-3}\sigma_{-4}\sigma_{11}\sigma_{10}\sigma_{12}\sigma_{9}\sigma_{11}\sigma_{13}\sigma_{8}\sigma_{10}\sigma_{12}\sigma_{14}\\
\hspace*{.45in}\sigma_{7}\sigma_{9}\sigma_{11}\sigma_{13}\sigma_{15}\sigma_{8}\sigma_{10}\sigma_{12}\sigma_{14}\sigma_{9}\sigma_{11}\sigma_{13}\sigma_{10}\sigma_{12}\sigma_{11}$}
\item{Braid on 13 strands.\\ $\sigma_L=\sigma_{10}\sigma_{11}\sigma_{10}\sigma_{12}\sigma_{11}\sigma_{10}\sigma_{-4}\sigma_{-5}\sigma_{-3}\sigma_{-6}\sigma_{-4}\sigma_{-2}\sigma_{-7}\sigma_{-5}\sigma_{-3}\sigma_{-1}\sigma_{-6}\sigma_{-4}\sigma_{-2}\sigma_{-5}\sigma_{-3}\sigma_{-4}$\\ $\sigma_R=\sigma_{-3}\sigma_{-2}\sigma_{-3}\sigma_{-1}\sigma_{-2}\sigma_{-3}\sigma_9\sigma_8\sigma_{10}\sigma_7\sigma_9\sigma_{11}\sigma_6\sigma_8\sigma_{10}\sigma_{12}\sigma_7\sigma_9\sigma_{11}\sigma_8\sigma_{10}\sigma_9$}
\item{Braid on 10 strands.\\ $\sigma_L= \sigma_{-3} \sigma_{-4}   \sigma_{ -2}    \sigma_{-5 }   \sigma_{-3 }   \sigma_{-1 }   \sigma_{-4}    \sigma_{-2}    \sigma_{-3}    \sigma_{-1}   \sigma_{ -2 }   \sigma_{-1}$ \\ $\sigma_R=     \sigma_{9}     \sigma_{8}    \sigma_{ 9}     \sigma_{7}     \sigma_{8}     \sigma_{6}     \sigma_{9}     \sigma_{7}     \sigma_{5}     \sigma_{8}     \sigma_{6}     \sigma_{7}$}
\item{Braid on 8 strands.\\ $\sigma_L=\sigma_{-1}\sigma_{-3}\sigma_{-4}\sigma_{-2}\sigma_{-3}\sigma_{-1}\sigma_{-2}$ \\ $\sigma_R=\sigma_{6}\sigma_{7}\sigma_{5}\sigma_{6}\sigma_{4}\sigma_{5}\sigma_{7}$}
\end{itemize}

\end{appendix}
\newpage

%

\end{document}